\documentclass[12pt]{article}
\pdfoutput=1
\usepackage[T1]{fontenc}
\usepackage{tikz,pgfplots}
\usepackage{multirow}
\usetikzlibrary{matrix}
\usepgflibrary{shapes.misc}
\usetikzlibrary{arrows,shadows}
\usepackage{graphicx}
\usepackage{bbold}
\usepackage[utf8]{inputenc}
\usepackage{amsmath}
\usepackage{mathtools}
\usepackage{amsfonts}
\usepackage{amssymb}
\usepackage{booktabs}
\usepackage[english]{babel}
\usepackage{setspace}
\usepackage{geometry}
\usepackage{xcolor}
\usepackage{color, colortbl}
\usepackage{tikz}
\newcommand{\bea}{\begin{eqnarray}}
\newcommand{\eea}{\end{eqnarray}}    
\newcommand{\be}{\begin{equation}}
\newcommand{\ee}{\end{equation}}

\begin{document}
\thispagestyle{empty}
$\,$

\vspace{32pt}
\begin{center}

\textbf{\Large  On the  systematic uncertainties in DUNE and their role in New Physics studies} 

\vspace{30pt}
D. Meloni$^a$
\vspace{16pt}

\textit{$^a$Dipartimento di Matematica e Fisica, 
Universit\`a di Roma Tre\\Via della Vasca Navale 84, 00146 Rome, Italy}\\
\vspace{16pt}

\texttt{davide.meloni@uniroma3.it}
\end{center} 
 \abstract
In the recent years experiments have established the existence of neutrino oscillations
and most of the oscillation parameters have been measured with a good accuracy.
The search for New Physics in neutrino oscillation will be an experimental concrete possibility in the next future. In this paper we investigate 
the ability of the DUNE facility to search for Non Standard Interaction (NSI) in neutrino propagation in matter, emphasizing 
the role of different assumptions on the shape and absolute normalization errors of both $\nu_e$ and $\nu_\mu$ signals.
We also study in detail the effects of NSI and systematics in the extraction of standard oscillation parameters. 

\section{Introduction}

The standard neutrino  mixing angles and mass differences  have been determined with
very good accuracy \cite{Esteban:2016qun}-\cite{deSalas:2017kay} apart from the yet unknown mass ordering, the sign of the
atmospheric mass squared difference, the octant of the atmospheric mixing angle well as
the CP-violating phase $\delta_{CP}$. Non-oscillation experiments have also provided important information 
on the neutrino cross sections on nuclei \cite{Benhar:2015wva}, which is triggering by itself an intense field of research. 
Excluding some hints for neutrino oscillation into 
additional sterile neutrino states \cite{Gariazzo:2018mwd}-\cite{Dentler:2018sju},  there are no experimental evidences that neutrinos possess non-standard
properties beyond those described by the Standard Model (SM). Thus, the presence  of new phenomena in the neutrino sector is still an open question 
which attracts a lot of attention in both theoretical and experimental communities, absorbed by the investigation of models able to produce small perturbations 
on top of the usual three-flavor oscillation and by the requirements of the experimental facilities needed to detect them.
In such an intriguing panorama, a crucial role is played by the systematic uncertainties which affect the sensitivity of a given experiment
to oscillation parameters and, even more seriously, to New Physics. For this reason, the relevant sources of uncertainty (like flux normalization, energy calibration, nuclear effects, detector
efficiency and so on) must be contemplated in any simulation of future detectors aiming at testing the standard paradigm of neutrino oscillation \cite{Ankowski:2016jdd}.
It is well known that correlated systematics can be reduced by using the unoscillated event distribution
at Near Detector to predict the distribution at the Far Detector; this strategy works very well for short-baseline experiments and has been used 
with great success in reactor experiments to measure $\theta_{13}$. Even though the Near to Far extrapolation at long baselines 
does not seem to provide  the same advantages in reducing the systematic uncertainties (because, for example, the absolute cross section is known with an accuracy of 10-20\%, 
the kinematics is more difficult to treat because of several interaction mechanisms and in appearance measurements initial and final states are different), it is 
nevertheless of crucial importance to estimate what the most important sources of systematics are and, consequently,  to understand the needs and requirements for the future generation of 
Near Detectors \cite{CENF}.
Since the final answer obviously depends on the chosen experimental facility and on what kind of new physics scenario we want to address, 
we start here this ambitious program checking which reasonable combinations of  {\it shape}
\footnote{With {\it shape uncertainty} we mean here the {\it uncorrelated bin-to-bin uncertainty}  in
the energy spectrum.} and   {\it absolute normalization}
uncertainties of both $\nu_e$ appearance (app) and $\nu_\mu$ disappearance (dis) signals can ensure the best performance in ameliorating the current bounds on 
the matter Non Standard Interaction 
(NSI) parameters (see \cite{Farzan:2017xzy} for a recent review) at the DUNE facility \cite{Acciarri:2015uup}. This will be the subject of 
Sect.\ref{sect:sys}.  At the same time, we update 
the existing bounds on NSI using the latest results from the global fits on standard oscillation parameters, Sect.\ref{sect:numres}, as well as 
the impact of marginalizing over NSI parameters in the estimate of the octant, CP and mass hierarchy determination sensitivities, Sect.\ref{sect:effect}.
Notice that a similar interesting analysis was presented in \cite{Ghosh:2017ged} where, however, only the overall systematics 
for signal and background normalization were allowed to vary. 
In this work we do not attempt to take simultaneously into account possible NSI affecting production and/or detection  
\cite{Grossman:1995wx} as we consider them as subleading effects compared to the matter effect modifications \cite{Blennow:2016etl}.

NSI affecting neutrino propagation are generated through four-fermion operators involving Dirac bilinears of neutrinos and of fermions in the Earth matter, which modify 
the neutrino evolution equation in the flavor basis according to  \cite{Roulet:1991sm}-\cite{Bergmann:1999rz}:
\begin{eqnarray} 
i \frac{d}{dt} \left( \begin{array}{c} 
                   \nu_e \\ \nu_\mu \\ \nu_\tau 
                   \end{array}  \right)
 = \left[ U_{PMNS} \left( \begin{array}{ccc}
                   0   & 0          & 0   \\
                   0   & \Delta_{21}  & 0  \\
                   0   & 0           &  \Delta_{31}  
                   \end{array} \right) U_{PMNS}^{\dagger} +  
                  A \left( \begin{array}{ccc}
            1 + \epsilon_{ee}     & \epsilon_{e\mu} & \epsilon_{e\tau} \\
            \epsilon_{e \mu }^*  & \epsilon_{\mu\mu}  & \epsilon_{\mu\tau} \\
            \epsilon_{e \tau}^* & \epsilon_{\mu \tau }^* & \epsilon_{\tau\tau} 
                   \end{array} 
                   \right) \right] ~
\left( \begin{array}{c} 
                   \nu_e \\ \nu_\mu \\ \nu_\tau 
                   \end{array}  \right)\, ,
\label{eq:matter}
\end{eqnarray}
where  $\Delta_{ij}=\Delta m^2_{ij}/2E$, $U_{PMNS}$ is the neutrino mixing matrix,
$A\equiv 2 \sqrt 2 G_F n_e$ and $\epsilon_{\alpha\beta}$ are effective parameters encoding the action of the four-fermion operators. Thus, 
beside the standard oscillation parameters, the parameter space is enriched by six more moduli $|\epsilon_{\alpha\beta}|$ 
and three more phases $\phi_{\alpha\beta}$, on which no constraints have been derived so far. 
A reduction of the number of independent moduli can be achieved subtracting to the whole Hamiltonian the matrix $\epsilon_{\tau\tau} \times \mathbb{1}$ and redefining
$\epsilon_{ee}- \epsilon_{\tau\tau}\to \epsilon_{ee}$ and $\epsilon_{\mu\mu}- \epsilon_{\tau\tau}\to \epsilon_{\mu\mu}$; in this case $\epsilon_{\tau\tau}$ 
is set to zero in our numerical simulations. 
The current 90\% Confidence Level (CL) bounds on the real part of the NSI parameters used throughout this paper are reported in Tab.\ref{tab:limits} 
and extracted from \cite{Coloma:2015kiu}. No constraints on the new CP phases are available so far.
\begin{table}[h!]
\begin{center}
\begin{tabular}{c | c }
\toprule
\midrule
  &   current limits \\ 
\midrule
 $\epsilon_{e e}$ & (-4, -2.62) $\oplus$ (0.33, 1.79) \\
 $|\epsilon_{e \mu}|$ & <0.36 \\
 $|\epsilon_{e \tau}|$ & <0.53 \\
 $\epsilon_{\mu\mu}$ & (-0.12, 0.11) \\
 $|\epsilon_{\mu\tau}|$ &  < 0.054\\
\midrule
\end{tabular}
\caption{\it Current constraints on the NSI parameters at 90\% CL obtained from a global fit to neutrino oscillation data \cite{Gonzalez-Garcia:2013usa}. 
All new CP violating phases are not constrained so far and thus have not been reported here.}
\label{tab:limits}
\end{center}
\end{table}

If $\theta_{23}$ is different from $\pi/4$ \cite{Kikuchi:2008vq} (but see also \cite{Kopp:2007ne}-\cite{Meloni:2009ia} for a detailed perturbative 
derivation of the transition probability formulae), the leading order dependence of 
$P(\nu_\mu \to \nu_e)$ and $P(\nu_\mu \to \nu_\mu)$ on the NSI parameters are reported in Tab.(\ref{tab:oscillation_exp}).
Here we use the short-hand notation $\epsilon_{\alpha\beta} = \epsilon$, with the meaning $\epsilon \lesssim {\cal O}(1)$ to identify a common order 
in the perturbative expansion of the probabilities.
\begin{table}[h!]
\begin{center}
\begin{tabular}{c  c  c  c  c  c }
\toprule
\midrule
 Probability &   $\epsilon_{ee}$ & $\epsilon_{e\mu}$ &  $\epsilon_{e\tau}$ & $\epsilon_{\mu\mu}$ & $\epsilon_{\mu\tau}$\\ 
\midrule
$P(\nu_\mu \to \nu_e)$ &  $\epsilon^3$ & $\epsilon^2$ & $\epsilon^2$ &  $\epsilon^3$ & $\epsilon^3$\\
 $P(\nu_\mu \to \nu_\mu)$ & $\epsilon^3$ & $\epsilon^2$ & $\epsilon^2$ &  $\epsilon$ & $\epsilon$ \\
\bottomrule
\end{tabular}
\caption{\it Leading order dependence of the oscillation probabilities $P(\nu_\mu \to \nu_e)$ and $P(\nu_\mu \to \nu_\mu)$ on the NSI parameters.}
\label{tab:oscillation_exp}
\end{center}
\end{table}
Thus, a change of systematics related to the $\nu_\mu \to \nu_e$ transition most probably will affect the determination of $\epsilon_{e\mu}$ and  $\epsilon_{e\tau}$, 
with scarse effects on the other parameters  $\epsilon_{\mu\mu}$ and $\epsilon_{\mu\tau}$, while a variation in the shape and absolute normalizations of the $\nu_\mu$ signal will reflect on a 
different sensitivity to $\epsilon_{\mu\mu}$ and $\epsilon_{\mu\tau}$.  For $\epsilon_{ee}$ the situation is a bit different; in fact, even though it is perturbatively suppressed
at ${\cal O}(\epsilon^3)$, nevertheless the current bounds allow it to assume ${\cal O}(1)$ values, thus making possible  a significant dependence on the systematic choice.
\section{Experimental setup and treatment of the systematics}
\label{sect:sys}
In this paper we focus on the DUNE experiment as described in \cite{Acciarri:2015uup}; we consider a 40 Kton liquid argon detector at a 
baseline $L=1300$ Km and on-axis with respect to 
the beam direction. The neutrino fluxes correspond to the Optimized Desing flux, from a proton beam energy of 80 GeV and a beam power of 1.07 MW. 
$3.5$ years of data taking 
are assumed in both neutrino and antineutrino modes, for a total exposure of 300 kt$\cdot$MW$\cdot$year for both $\nu_e$ appearance and $\nu_\mu$ 
disappearance. 
No near detector is assumed in our numerical simulations. At the far detector the $\nu_\mu$ disappearance sample
is composed of $\nu_\mu$ CC interactions, with main 
backgrounds from neutral current (NC) interactions where charged pions are misidentified as muons, and $\nu_{\tau}+ \bar \nu_{\tau}$ CC interactions in which the produced $\tau$s
decay to a muon and two neutrinos.
The $\nu_e$ appearance sample is composed of $\nu_e$ CC interactions from  $\nu_\mu  \to \nu_e$ oscillation
and the relevant backgrounds originate from intrinsic  $\nu_e + \bar \nu_e$ beam contamination, NC and
$\nu_\mu + \bar \nu_\mu$ CC interactions in which a photon from electromagnetic neutral pion decays is
misidentified as an electron, and from $\nu_{\tau} + \bar \nu_{\tau}$ interactions in which the resulting $\tau$
decays to an electron. It should be noticed that in both neutrino and antineutrino running modes we consider the sum $\nu_\mu \to \nu_e \oplus \bar \nu_\mu \to \bar \nu_e$ 
as signal appearance events because the information coming from the wrong-sign events can be of some relevance. According to \cite{Alion:2016uaj}, 
the same approach is also used for the disappearance channels,
for which the sum $\nu_\mu \to \nu_\mu \oplus \bar \nu_\mu \to \bar \nu_\mu$ is considered.

The Liquid Argon Time Projection Chamber (LArTPC) performance parameters used to get the sensitivity plots discussed in this paper 
have been generated using the DUNE Fast Monte Carlo simulation \cite{Adams:2013qkq}
and translated into GLoBES files \cite{Huber:2004ka,Huber:2007ji} made publicly available as ancillary files in \cite{Alion:2016uaj}. Beside cross section files describing neutrino charged and neutral current interactions
in Argon (generated using GENIE 2.8.4), the DUNE Collaboration also provided true-to-reconstructed smearing matrices (not modified in this work) as well as selection efficiencies as a function of 
bin energies for the various signals and backgrounds.

Our implementation of the $\chi^2$  and the method of determining the confidence regions is based on the pull method 
\cite{Huber:2002mx,Fogli:2002pt,Ankowski:2016jdd}
and represents the standard implementation of systematic uncertainties in GLoBES.
The $\chi^2$ is obtained after the minimization over the nuisance parameters $\vec \xi$ \cite{Ankowski:2016jdd}.
For each transition channel $c$ and energy bin $i$ we use a Poissonian $\chi^2$ of the form:
\begin{equation}
  \chi_c^2 = \sum_i  2 \bigg( F_{c,i}(\vec{\theta}, \vec{\xi}) - O_{c,i}
                            + O_{c,i} \ln \frac{O_{c,i}}{F_{c,i}(\vec{\theta}, \vec{\xi}))} \bigg) \,,
\label{equ:chirule}
\end{equation}
where $F_{c,i}(\vec{\theta}, \vec{\xi})$ is the predicted number of
events in the $i$-th energy bin for a given channel $c$, for a set of
oscillation parameters $\vec{\theta}$ and nuisance parameters $\vec{\xi}$.  $O_{c,i}$, instead, is the observed
event rate, that is the one corresponding to  assumed true values of the 
oscillation parameters.  Both $F_{c,i}$ and $O_{c,i}$ receive
contributions from different sources $s$, that tipically include signal and background rates specified by $R_{c,s,i}(\vec{\theta})$, 
so that for example: 
\begin{equation}
  F_{c,i}(\vec{\theta}, \vec{\xi}) = \sum_s \left(1 + a_{c,s}(\vec{\xi}) \right) R_{c,s,i}(\vec{\theta}) \,.
\end{equation}
The auxiliary parameters $a_{c,s}$ have the form $a_{c,s} \equiv \sum_k w_{c,s,k} \, \xi_k \,,
$ in which the coefficients $w_{c,s,k}$ can assume the values one or zero depending on whether 
a particular nuisance parameter $\xi_k$ affects the
contribution from the source $s$ to channel $c$   or not, respectively.
 
Thus, the total $\chi^2$ is given by:
\[
\chi^2=\min_\xi\left\{\sum_{\,c}\chi^2_{c}
+\sum_i\left(\frac{\xi_{\phi,\,i}}{\sigma_\phi}\right)^2+
\left(\frac{\xi_N}{\sigma_N}\right)^2\right\},
\]
where the last two contributions are the pull terms associated with the shape and the overall signal normalization, respectively. The $\xi_{\phi,\,i}$ shape parameters 
are bin-to-bin uncorrelated whereas the normalization parameter $\xi_N$ is fully correlated between different energy bins $i$. 

In \cite{Alion:2016uaj} only overall normalizations for signal and backgrounds have been taken into account; for $\nu_e$ and  $\bar \nu_e$  signal modes they are fixed to 2\% each, 
while a 5\% is assumed for $\nu_\mu$ and $\bar \nu_\mu$ signals. For the backgrounds the normalization uncertainties range from 5\% (for intrinsic $\nu_e$ and misidentified $\nu_\mu$) to 20\% 
(for misidentified $\tau$) with a 10\% assumed for the neutral currents. While the background  normalization uncertainties are not changed in our numerical simulations, 
we consider several options for the signal normalization uncertainties and, in addition, we also take into account various level of normalizations.
In order to maintain the possible combinations of shape and normalization errors to a reasonable level, we decided to split the systematic uncertainties in two different classes: those in which only the $\nu_e$ signal
uncertainties are changed (and the ones related to $\mu$'s are fixed, case-$I$) and, viceversa,  those in which only the $\nu_\mu$ signal
uncertainties are changed (case-$II$). Thus the following combinations for case-$I$ are considered:
\[
  \underline{{\rm  case}-I}= 
                \begin{array}{ll}
                  {\rm absolute\; normalization:} \begin{cases}
               \nu_\mu{\rm\; dis = 5\%}\\
                \nu_e{\rm\; app = 2.5\%,  5\%}
            \end{cases}\\
                   \\
                  {\rm shape\; normalization:}\begin{cases}
               \nu_\mu{\rm\; dis = 5\%}\\
                \nu_e{\rm\; app = 2\%,  7\%}
            \end{cases}\\
                \end{array} \,,       
\]
which give rise to four different scenarios: 
\begin{itemize}
 \item $A$, very optimistic, where the pair (shape, absolute) = (2\%, 2.5\%);
 \item $B_1$, (shape, absolute) = (2\%, 5\%);
 \item $B_2$, (shape, absolute) = (7\%, 2.5\%);
 \item $C$, very pessimistic, with (shape, absolute) = (7\%, 5\%).
\end{itemize}
Analogously, for  case-$II$ we have:
\[
  \underline{{\rm  case}-II}= 
                \begin{array}{ll}
                  {\rm absolute\; normalization:} \begin{cases}
               \nu_\mu{\rm\; dis = 2\%, 5\%}\\
                \nu_e{\rm\; app = 2.5\%}
            \end{cases}\\
                   \\
                  {\rm shape\; normalization:}\begin{cases}
               \nu_\mu{\rm\; dis =  2\%,  7\%}\\
                \nu_e{\rm\; app = 2.5\%}
            \end{cases}\\
                \end{array} \,,       
\]
which gives rise to the four scenarios reported below (for them, we use the subscript $\mu$): 
\begin{itemize}
 \item $A_\mu$, (shape, absolute) = (2\%, 2.5\%);
 \item $B_{1\mu}$, (shape, absolute) = (2\%, 5\%);
 \item $B_{2\mu}$, (shape, absolute) = (7\%, 2.5\%);
 \item $C_\mu$, (shape, absolute) = (7\%, 5\%).
\end{itemize}
It is worth mentioning that systematic uncertainties between 2\% and 5\%  have also been assumed in \cite{Acciarri:2015uup} whereas the pessimistic 7\% has been introduced by us (but also 
taken into account in \cite{Ghosh:2017ged}) to contemplate
more conservative estimates on the physics reach of DUNE. For the sake of simplicity, while retaining the nominal binning for the spectrum,
we adopt wider intervals 
for the systematic uncertainties: [0.5,1] - [1,2] - [2,3] -[3,5] -[5,8 or 20] GeV.

\section{Numerical results}   
\label{sect:numres}
In this section we discuss in details how the bounds on the NSI parameters change if different assumptions on systematics are made according to the 
discussion of the previous section. In deriving the CL intervals for a given $\epsilon_{\alpha\beta}$, we marginalize over all
other standard and non-standard parameters: for  the NSI's, the  phases are left free in the whole 
$[0,2\pi)$ interval while the moduli are marginalized taking into account the 90\% CL ranges quoted in Tab.(\ref{tab:limits}).
Instead, for the central values and relative uncertainties of the standard mixing angles and mass differences, distinguished for Normal Ordering (NO) and 
Inverted Ordering (IO) whenever necessary,
we adopt the latest results in \cite{Esteban:2016qun}, see Tab.(\ref{bestfit}), but for  the leptonic CP 
phase $\delta_{CP}$ which, as the other phases, is left free in $[0,2\pi)$ \footnote{Since the main goal of the paper is to 
study the effects of the systematics on the study of the NSI scenario more than deriving new bounds on the 
NSI parameters, we adopted here the simple approach to take external priors on the standard mixing parameters.}. For the matter density parameter $A$ we used (and kept fixed) the standard constant value 
$A = 10.64\cdot 10^{-14}$ eV$^{-1}$ \footnote{As shown in \cite{dunemeeting}, bounds obtained 
with all the parameters free to vary are not importantly affected by the Earth
profile.}. All numerical results of this section have been obtained using a modified version of GLoBES which includes non-standard interactions 
affecting the  propagation  processes for active neutrinos \cite{Kopp:2006wp}. 
\begin{table}
\begin{center}
\begin{tabular}{c  c  c}
\toprule
\midrule
parameter & central value ($^\circ$)& relative uncertainty \\
\midrule
$\theta_{12}$ & 33.62 & 2.3\% \\ 
$\theta_{23}$ (NO) & 47.2  & 4.0\% \\ 
$\theta_{23}$ (IO) & 48.1  & 3.6\% \\ 
$\theta_{13}$ & 8.54  & 1.8\% \\ 
$\Delta m^2_{21}$ & 7.4$\times10^{-5}$~eV$^2$ & 2.8\% \\ 
$\Delta m^2_{31}$ (NO) & 2.49$\times10^{-3}$~eV$^2$ &  1.3\% \\ 
$\Delta m^2_{31}$ (IO) & -2.46$\times10^{-3}$~eV$^2$ &  1.3\% \\
\midrule
\bottomrule
\end{tabular}
\caption{\it Central values and relative uncertainties of the standard mixing parameters extracted from \cite{Esteban:2016qun}. 
 As in \cite{Acciarri:2015uup}, for non-Gaussian parameters the relative 
 uncertainty is computed using 1/6 of the 3$\sigma$ allowed range.}
\label{bestfit}
\end{center}
\end{table}
The event rates for signal and background in both neutrino and antineutrino modes for $\epsilon_{\alpha\beta}=0$ are reported in Tab.(\ref{rates}), 
where efficiencies have been taken into account. For $\nu_e$ appearance we take neutrino energies 
in the interval $E_\nu \in [0.5, 8]$ GeV while for $\nu_\mu$ disappearance we consider $E_\nu \in [0.5,20]$ GeV. The adopted bin sizes are the same as in \cite{Alion:2016uaj} 
\footnote{A slight increase or reduction in the number of bins 
should not have a significant impact on the sensitivity reach of DUNE \cite{elizabeth}.}.
\begin{table}[h!]
\begin{center}
\renewcommand\arraystretch{1.2}
\begin{tabular}{c || c || c | c |c| c}
\toprule
\midrule
 & \bf signal& \multicolumn{4}{c}{\bf backgrounds} \\
\midrule
& &intrinsic $\nu_e$  & mis $\nu_\mu$  & mis $\nu_\tau$ & NC \\ 
 & $\nu_\mu \to \nu_e \oplus \bar \nu_\mu \to \bar \nu_e$  &  &   & &  \\ 
\multirow{2}*{\bf neutrino mode} & 1161 $\oplus$ 12.9   & 292.0 & 2.8 & 20.1 & 26.0 \\ 
 & $\nu_\mu \to \nu_\mu \oplus \bar \nu_\mu \to \bar \nu_\mu$ & &  & &  \\
 &7630 $\oplus$ 504.4 & & &30.2 & 76.2\\ \hline
  & $\bar \nu_\mu \to \bar \nu_e \oplus  \nu_\mu \to  \nu_e$  & &   & &  \\
\multirow{2}*{\bf antineutrino mode} & 201 $\oplus$ 62   & 173.1 & 1.6 & 11.7 & 12.7 \\ 
 & $\bar \nu_\mu \to \bar \nu_\mu \oplus \nu_\mu \to \nu_\mu$ & &  & &  \\
 &2568 $\oplus$ 1500 & & &18.8 & 40.7\\
\midrule 
\bottomrule
\end{tabular}
\caption{\it Total number of signal and background events for both neutrino and antineutrino modes, 
computed for the oscillation parameters fixed as in Tab.(\ref{bestfit}) but $\delta_{CP}$, which is assumed to be vanishing. All NSI parameters are set to 
zero.
Notice that for the signal events we have considered both CP conjugate channels.}
\label{rates}
\end{center}
\end{table}
 
\subsection{Results for case-$I$} 
\begin{figure}[h!]
\begin{center}
  \includegraphics[scale=0.49]{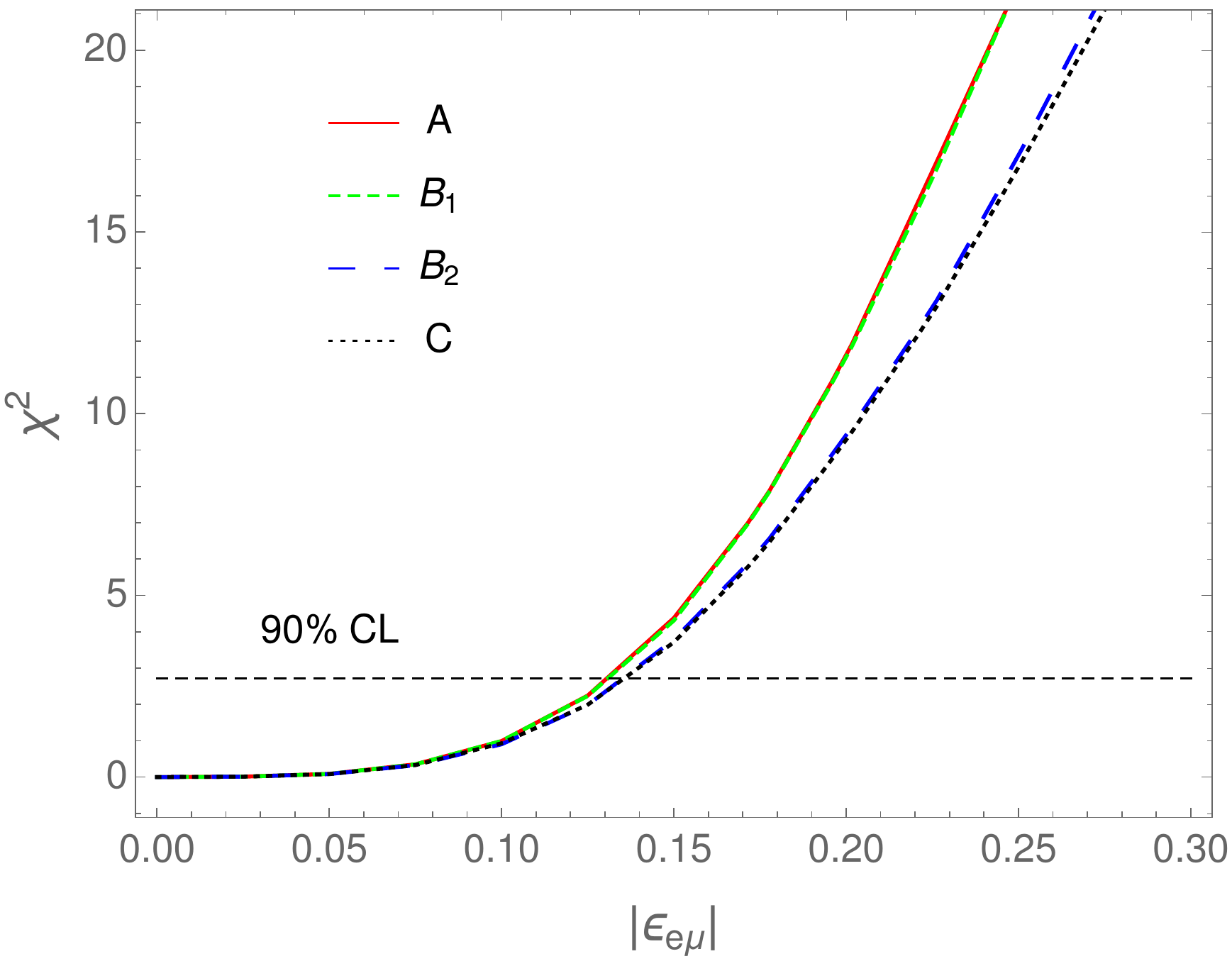} \includegraphics[scale=0.48]{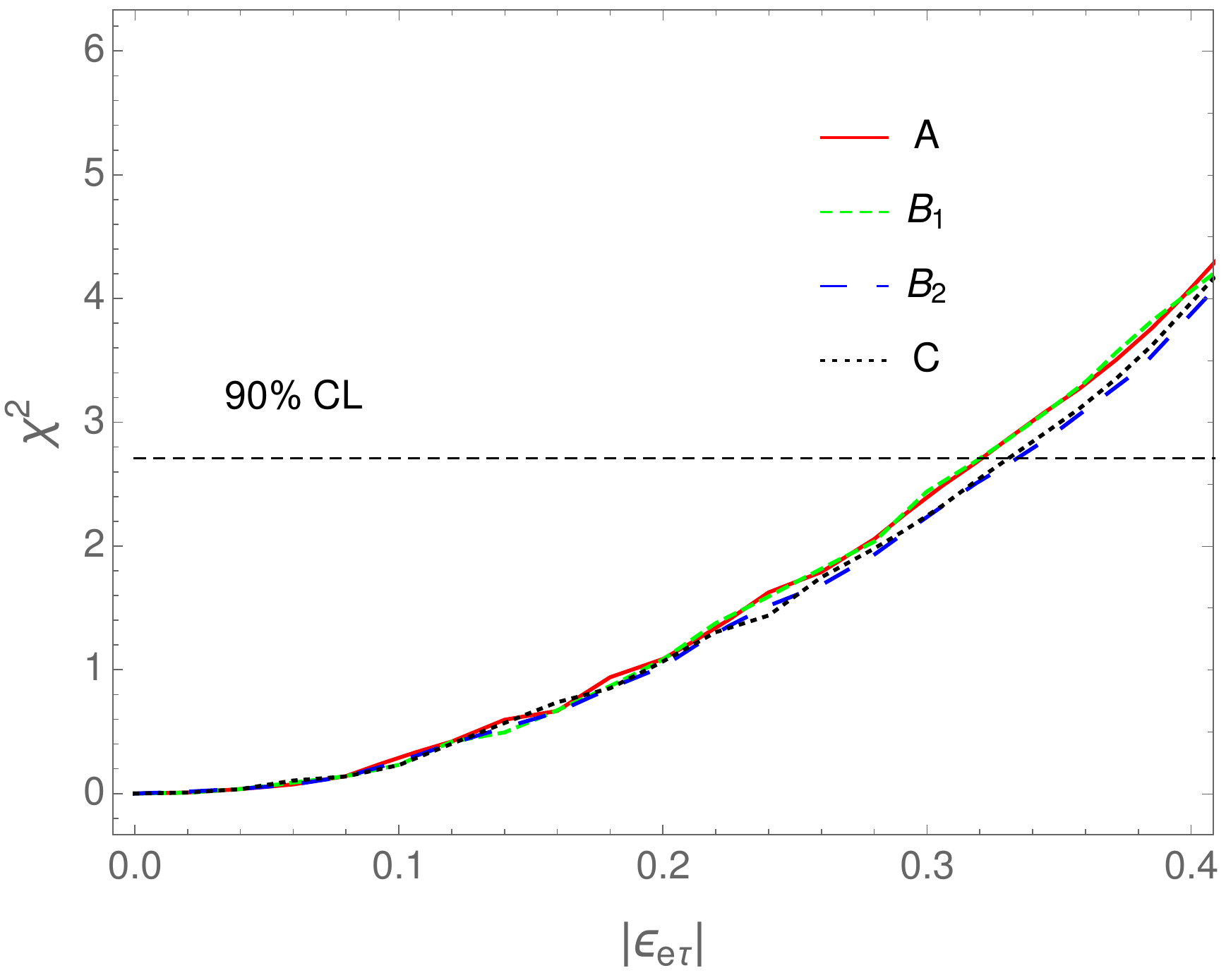}
\caption{\it \label{fig:chi2} Case-$I$: $\chi^2$ function for the NSI parameters $|\epsilon_{e\mu}|$ (left panel) and $|\epsilon_{e\tau}|$ (right panel). The red/solid, green/dashed, blue/long-dashed 
and black/dotted lines 
refer to the four different assumptions on systematics $A$, $B_1$, $B_2$ and $C$ on the $\nu_e$ signal.
Previous constraints on NSI parameters given in Tab.(\ref{tab:limits}) have been considered in this figure. The horizontal line indicates the 90\% CL cut
on the $\chi^2$ function.
The parameters not shown are marginalized over, see text for details. }
\end{center}
\end{figure}

According to  the discussion after Tab.(\ref{tab:limits}), we start  considering the bounds at the 
90\% CL (1 degree of freedom) that DUNE can set on the parameters $|\epsilon_{e\mu}|$ (left panel of  Fig.(\ref{fig:chi2})) and $|\epsilon_{e\tau}|$ (right panel), for the cases $A$ (red, solid line), 
$B_1$ (green, dashed line), 
$B_2$ (blue, long-dashed line) and $C$ (black, dotted line).
Several comments are in order:
\begin{itemize}
\item  it is clear that the behavior of the $\chi^2$ function is quite similar for the pairs ($A$, $B_1$) and ($B_2$, $C$), signalizing that the 
shape uncertainty (which is the same within the pairs) is by far more relevant than the overall signal normalization. In fact, even doubling it from 2.5\% (cases $A$ and $B_2$) 
to 5\% (cases $B_1$ and $C$) we hardly see any significant difference. In fact, this has to be expected since the disappearance distribution 
fixes the normalisation of both muon neutrino and antineutrino event samples;
\item in the most favorable cases ($A$, $B_1$), DUNE can set a bound on $|\epsilon_{e\mu}|$ at the 90\% CL around $b_{\epsilon_{e\mu}}=0.130$ while  in the worst cases ($B_2$, $C$)
this is deteriorated to  $b_{\epsilon_{e\mu}}=0.135$, thus a $\sim$4\% less stringent bound 
(not a significant difference given the simple treatment of the systematics adopted in this paper);
\item for $|\epsilon_{e\tau}|$ the bounds at the 90\% CL are $b_{\epsilon_{e\tau}}\sim 0.32$ and $b_{\epsilon_{e\tau}}\sim 0.33$  
for the cases  ($A$, $B_1$) and ($B_2$, $C$), respectively, which entails a worsening by $\sim$3\%.
\end{itemize}
We have also checked that a further increase of the absolute normalization systematics to 7\% does not produce 
relevant changes in the obtainable bounds on  $|\epsilon_{e\mu}|$ and $|\epsilon_{e\tau}|$. 

\begin{figure}[h!]
\begin{center}
  \includegraphics[scale=0.32]{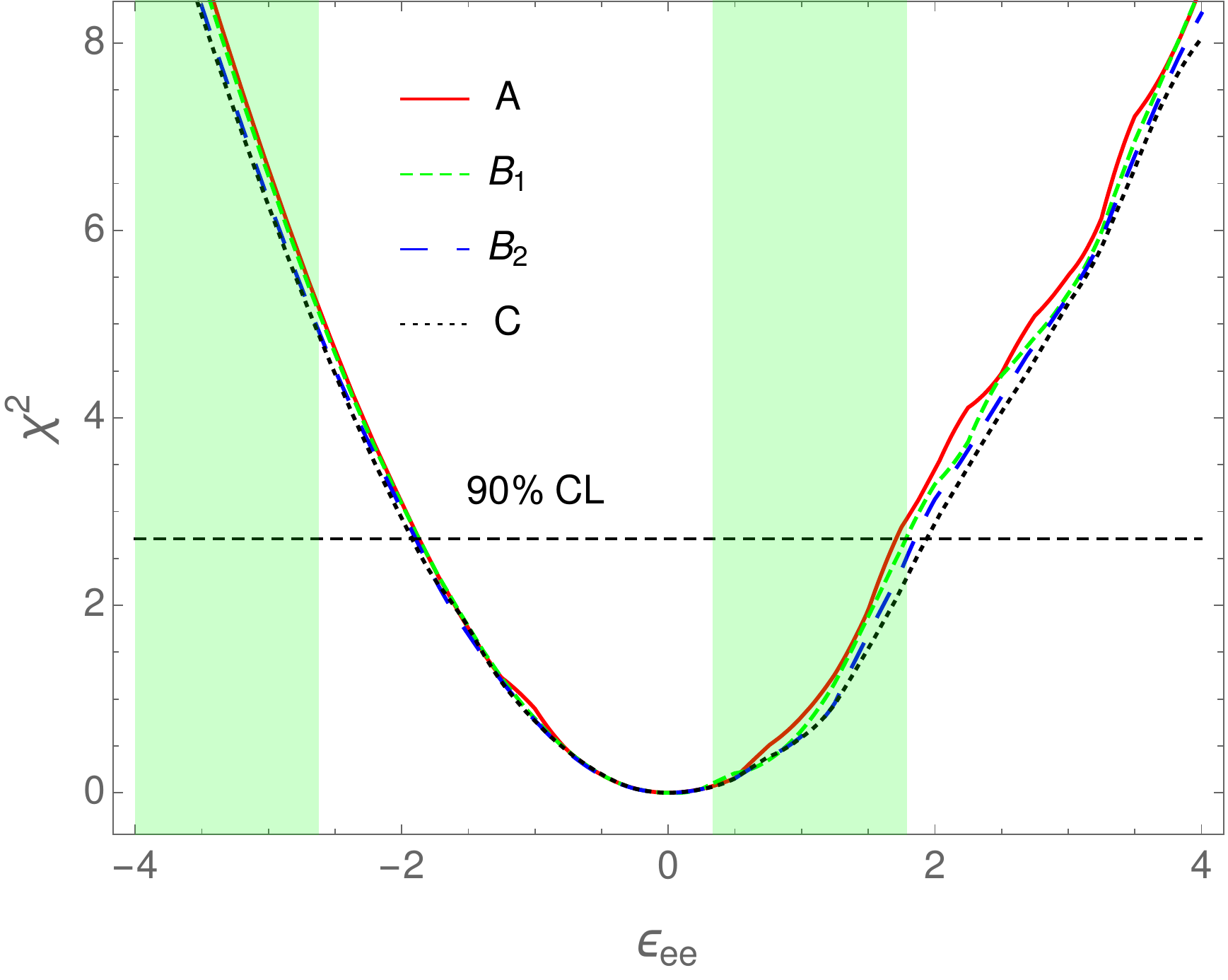} \includegraphics[scale=0.32]{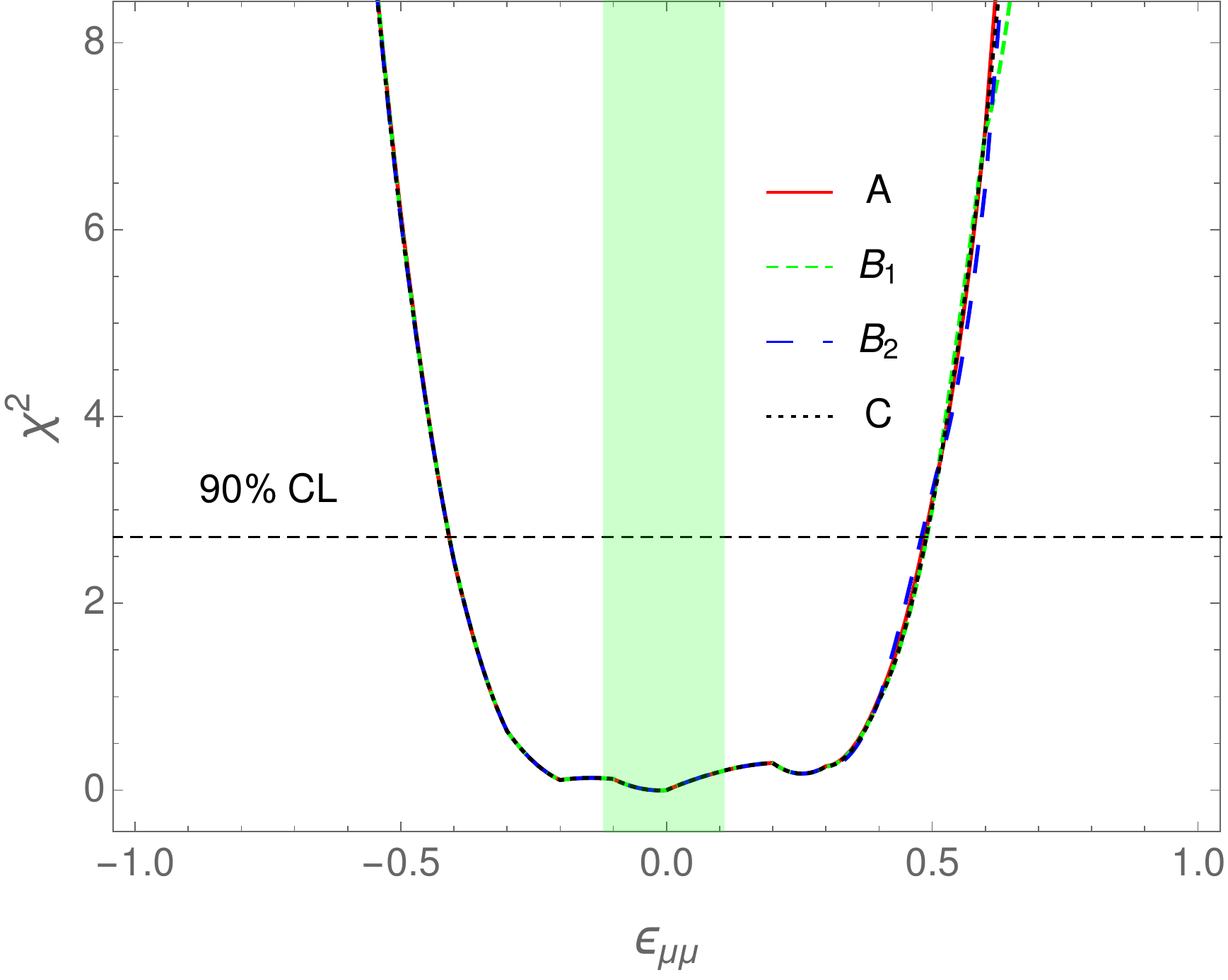} \includegraphics[scale=0.32]{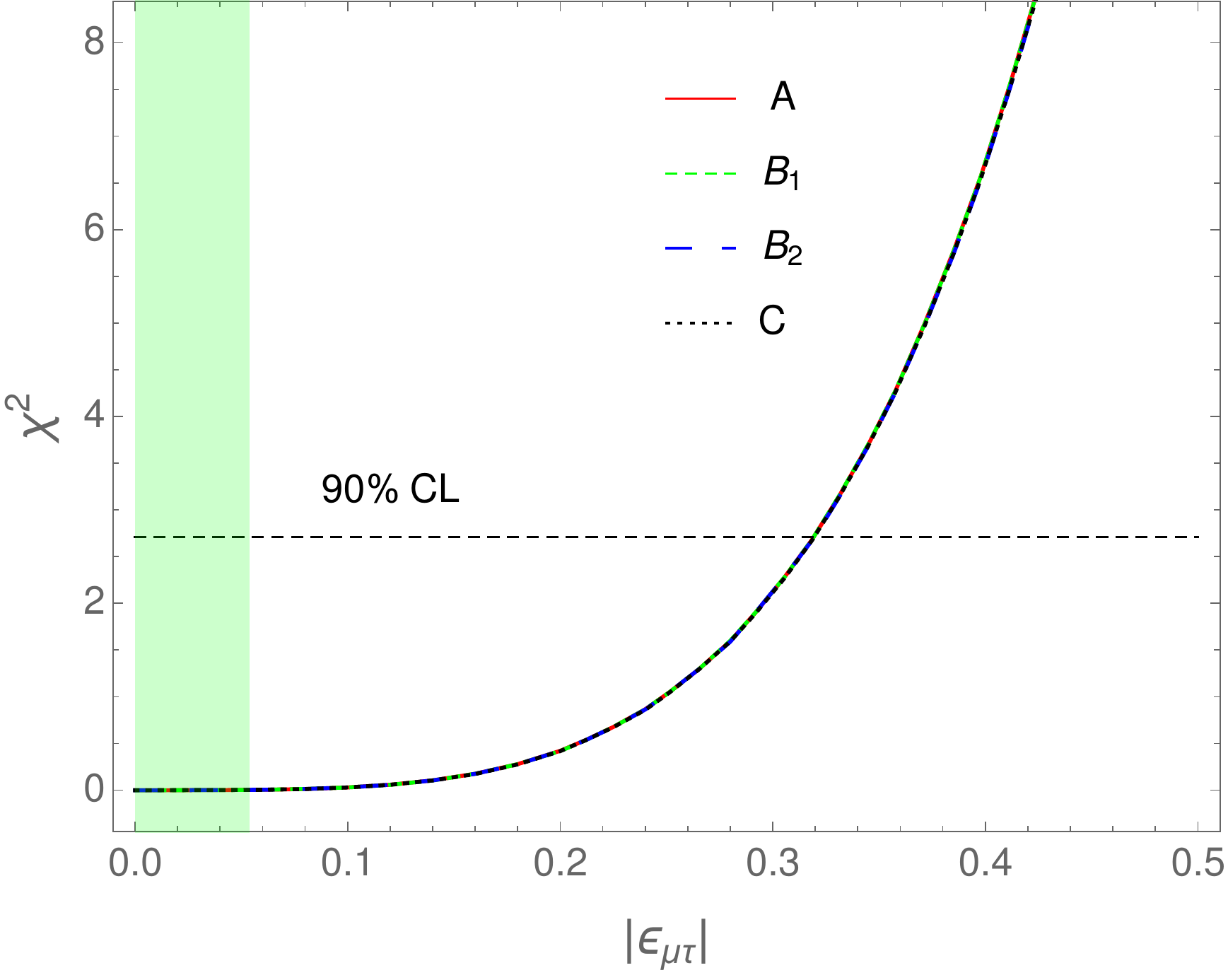}
\caption{\it \label{fig:chi22} The same as Fig.(\ref{fig:chi2}) but for the NSI parameters $\epsilon_{ee}$ (left panel), $\epsilon_{\mu\mu}$ 
(central panel) and $|\epsilon_{\mu\tau}|$ (right panel). Vertical (green) bands represent the 90\% CL allowed regions reported in Tab.(\ref{tab:limits}).}
\end{center}
\end{figure}

As already anticipated, for the other parameters  $\epsilon_{\mu\mu}$ and $|\epsilon_{\mu\tau}|$ there is no important dependence on the assumption 
on the systematic errors, as we can see in the central and right panels of Fig.(\ref{fig:chi22}).
However,  some dependence on systematics is visible for  $\epsilon_{ee}$ (left panel of Fig.(\ref{fig:chi22})): considering the extreme cases $A$ and $C$,  the allowed 90\% CL 
interval for case $A$ is about 7\% smaller than for case $C$.

For the sake of completeness, we summarize in Tab.(\ref{tab:summary}) the 90\% CL bounds on the five real parts of the $\epsilon_{\alpha\beta}$ parameters
\footnote{We verified that no significant constraints can be inferred on the new CP phases $\phi_{e\mu,e\tau,\mu\tau}$.}. 
\begin{table}
\begin{center}
\renewcommand\arraystretch{1.5}
\begin{tabular}{c  c  c  c  c}
\toprule
\midrule
\bf parameter & $A$& $B_1$ & $B_2$ & $C$ \\
\midrule
$ \epsilon_{ee}$ & [-1.87, 1.70] & [-1.87, 1.78] & [-1.90, 1.85] & [-1.92, 1.93] \\ 
$|\epsilon_{e\mu}|$   & [0, 0.130]& [0, 0.131]& [0, 0.135]& [0, 0.135]\\ 
$|\epsilon_{e\tau}|$   & [0, 0.321] &[0, 0.320] &[0, 0.334] &[0, 0.331] \\ 
$\epsilon_{\mu\mu}$   & [-0.41, 0.49] & [-0.41, 0.49]& [-0.41, 0.48]& [-0.41, 0.49]\\ 
$|\epsilon_{\mu\tau}|$  & [0, 0.320]& [0, 0.320]& [0, 0.320]& [0, 0.320]\\ 
\midrule
\bottomrule
\end{tabular}
\caption{\it  
Case-$I$: summary of the 90 \% CL bounds that DUNE can set on the real parts of the NSI parameters  as the systematic uncertainties vary from $A$ to $C$.
}
\label{tab:summary}
\end{center}
\end{table}
Comparing with the limits quoted in Tab.(\ref{tab:limits}), we observe that a strong improvement can be set on the bounds of $|\epsilon_{e\mu}|$, 
which passes from $b_{\epsilon_{e\mu}}< 0.36$ to $b_{\epsilon_{e\mu}}\lesssim 0.13$  for almost every choice of the systematics. 
For $|\epsilon_{e\tau}|$, instead, in all cases a reduction of the upper bound by roughly $\sim$40\% can be obtained.
As expected, 
the other off-diagonal parameters whose dependence in the oscillation probability $P(\nu_\mu \to \nu_e)$ is subleading, do not benefit of any improvements of the 
systematics, even in the more aggressive case $A$.
A separate discussion should be devoted to $\epsilon_{ee}$; in fact, while the current limit consists of two separate islands above and 
below $\epsilon_{ee}=0$, DUNE alone  will be able to exclude negative $\epsilon_{ee}$ but not those positive values within the current bounds 
\cite{Coloma:2015kiu}. 

\subsection{Results for case-$II$}
\begin{figure}[h!]
\begin{center}
  \includegraphics[scale=0.49]{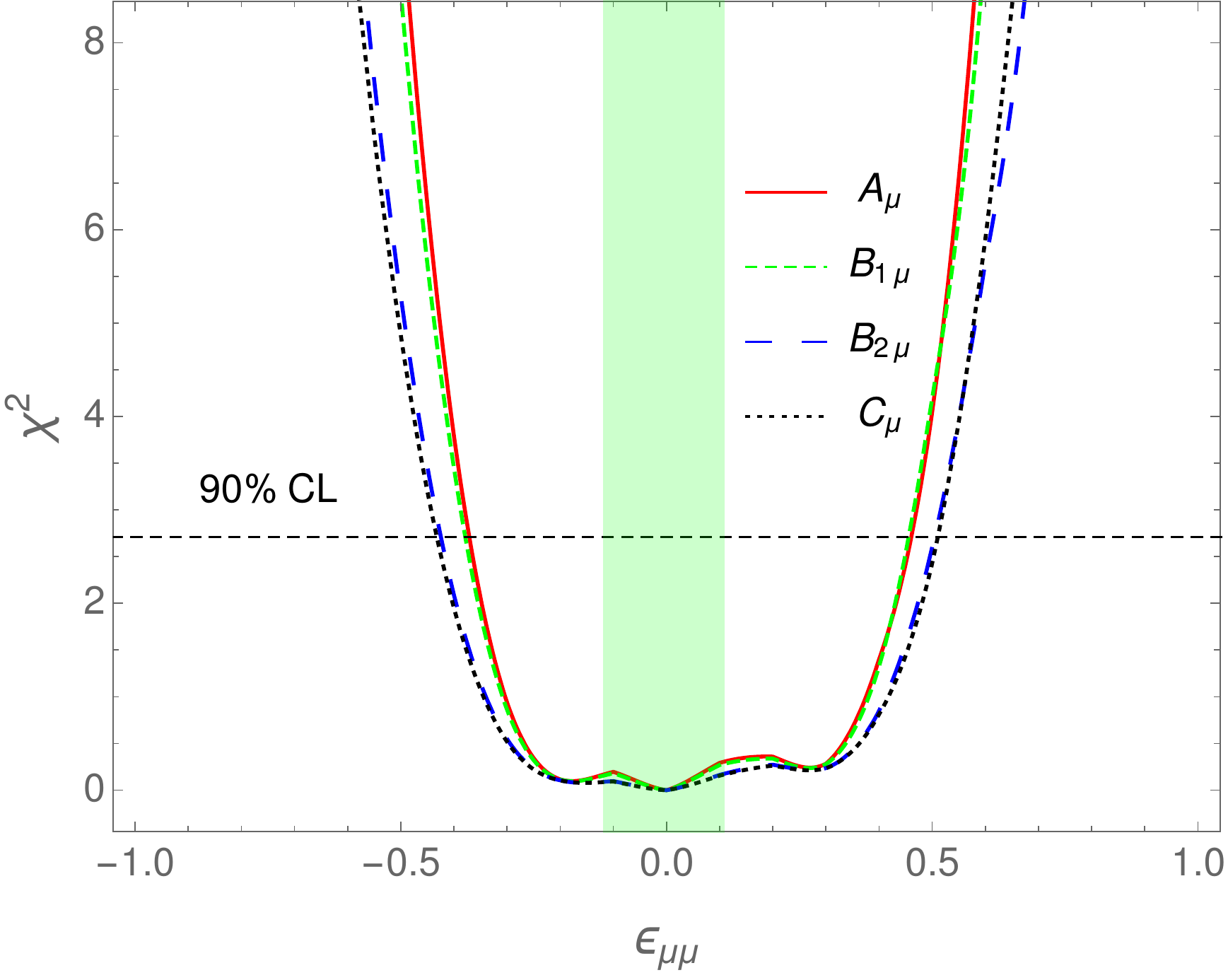} \includegraphics[scale=0.49]{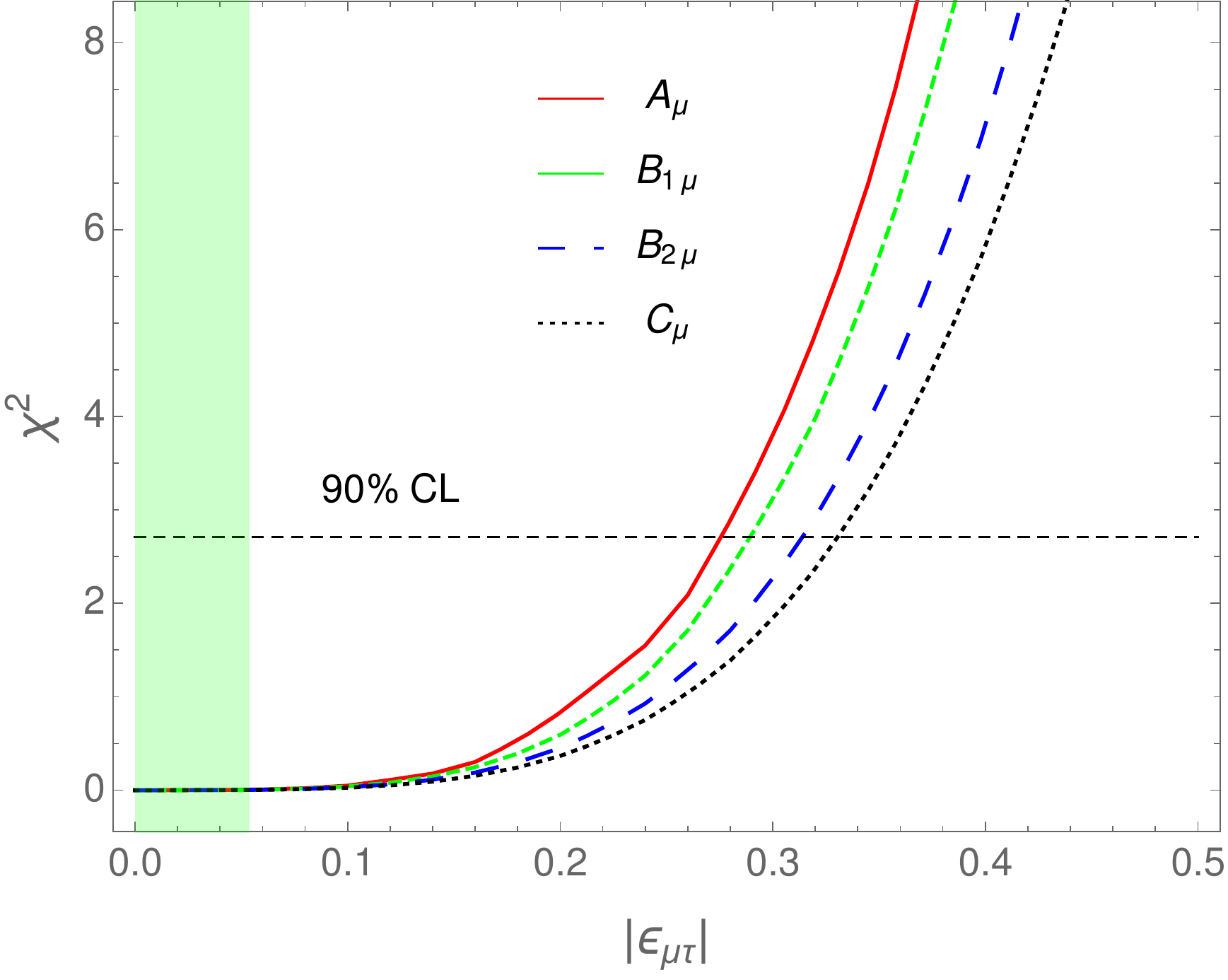}
\caption{\it \label{fig:chi2mu} Case-$II$: $\chi^2$ function for the NSI parameters $\epsilon_{\mu\mu}$ (left panel) and $|\epsilon_{\mu\tau}|$ (right panel). 
The red/solid, green/dashed, blue/long-dashed 
and black/dotted lines 
refer to four different assumptions $A_\mu$, $B_{1\mu}$, $B_{2\mu}$ and $C_\mu$ on the $\nu_\mu$ signal systematics studied in this paper.
Previous constraints on NSI parameters given in Tab.(\ref{tab:limits}) have been considered in this figure. The horizontal line indicates the 90\% CL cut
on the $\chi^2$ function.
The parameters not shown are marginalized over. Vertical (green) bands represent the 90\% CL allowed regions summarized in Tab.(\ref{tab:limits}).}
\end{center}
\end{figure}

\begin{figure}[h!]
\begin{center}
  \includegraphics[scale=0.32]{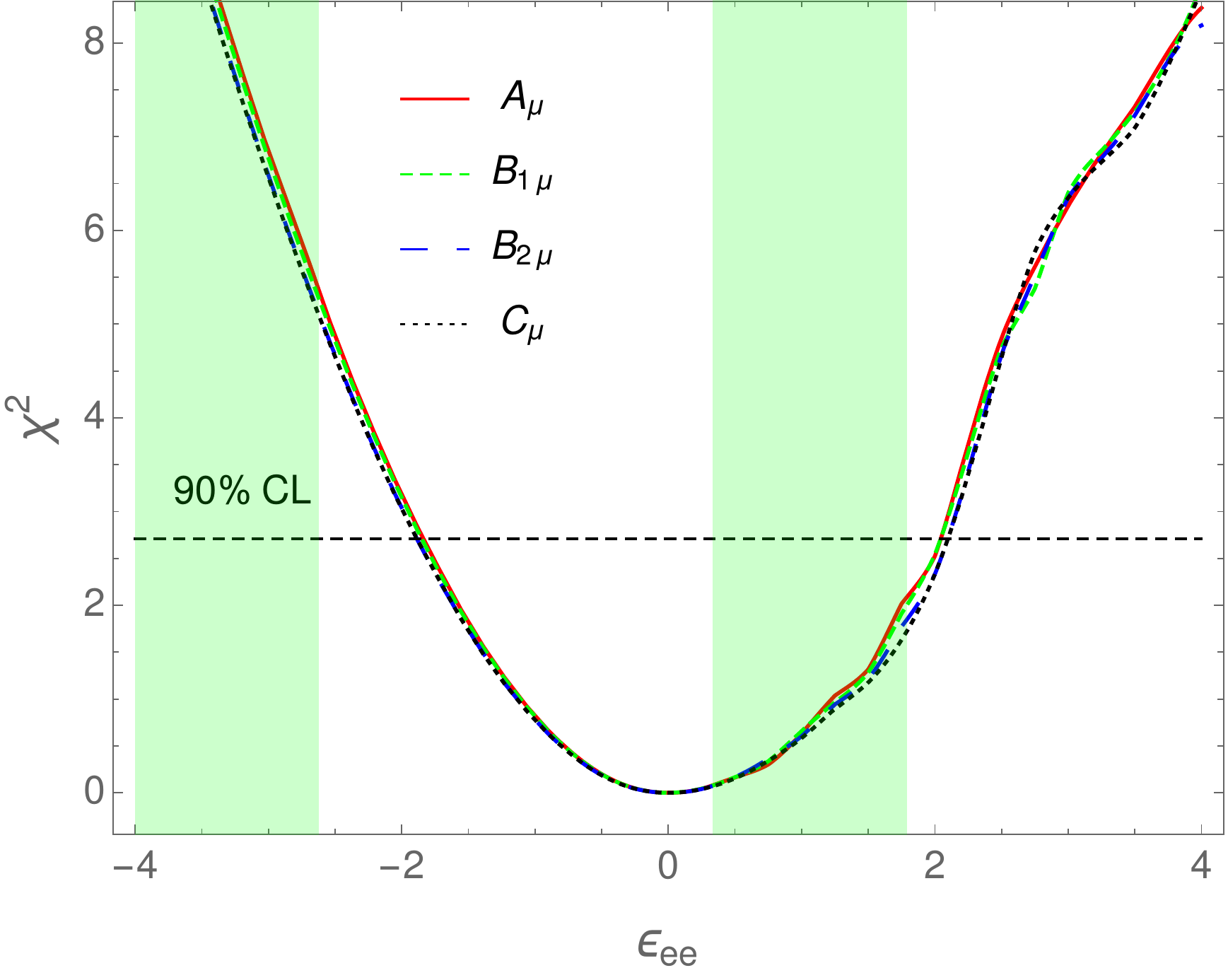} \includegraphics[scale=0.32]{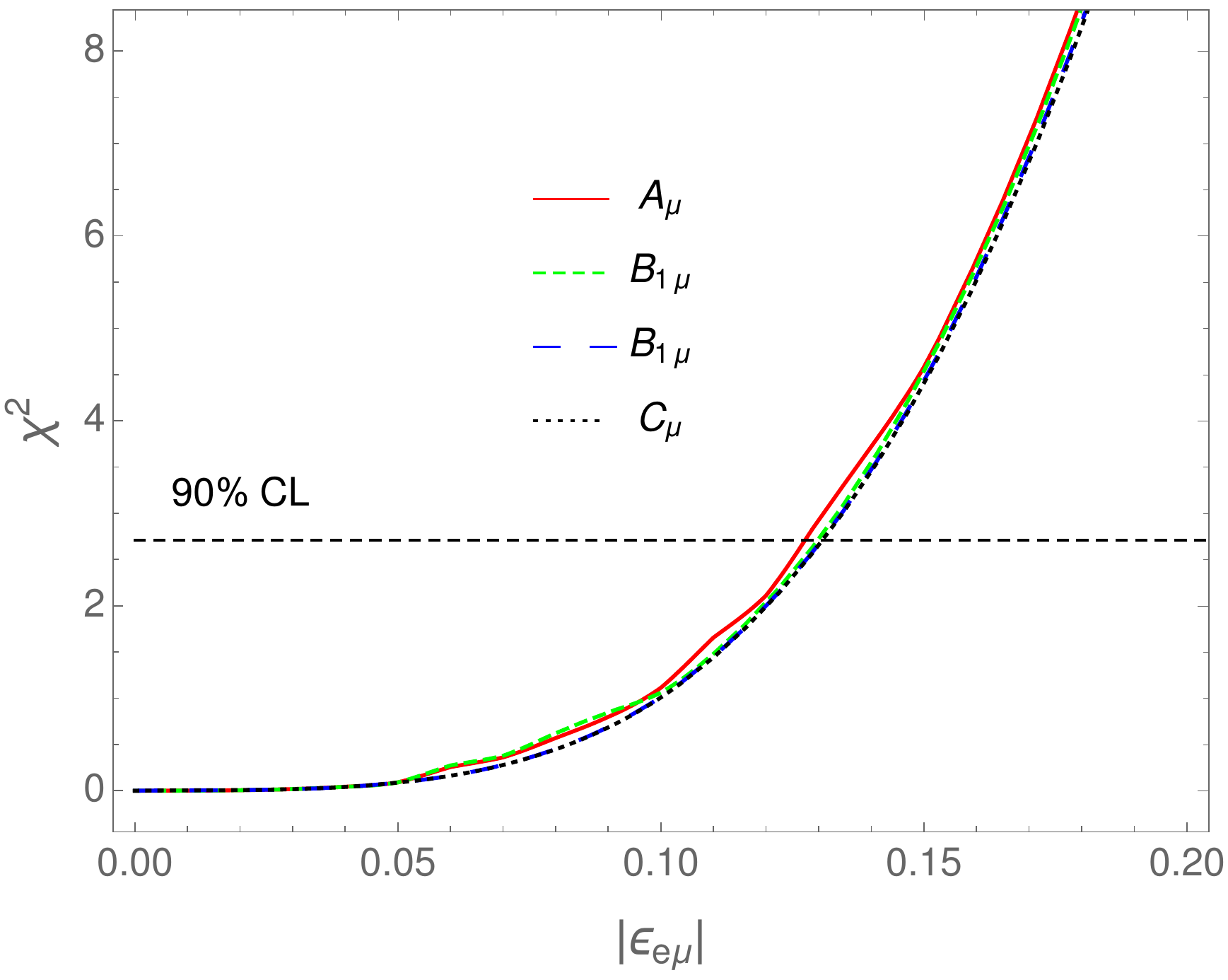} \includegraphics[scale=0.32]{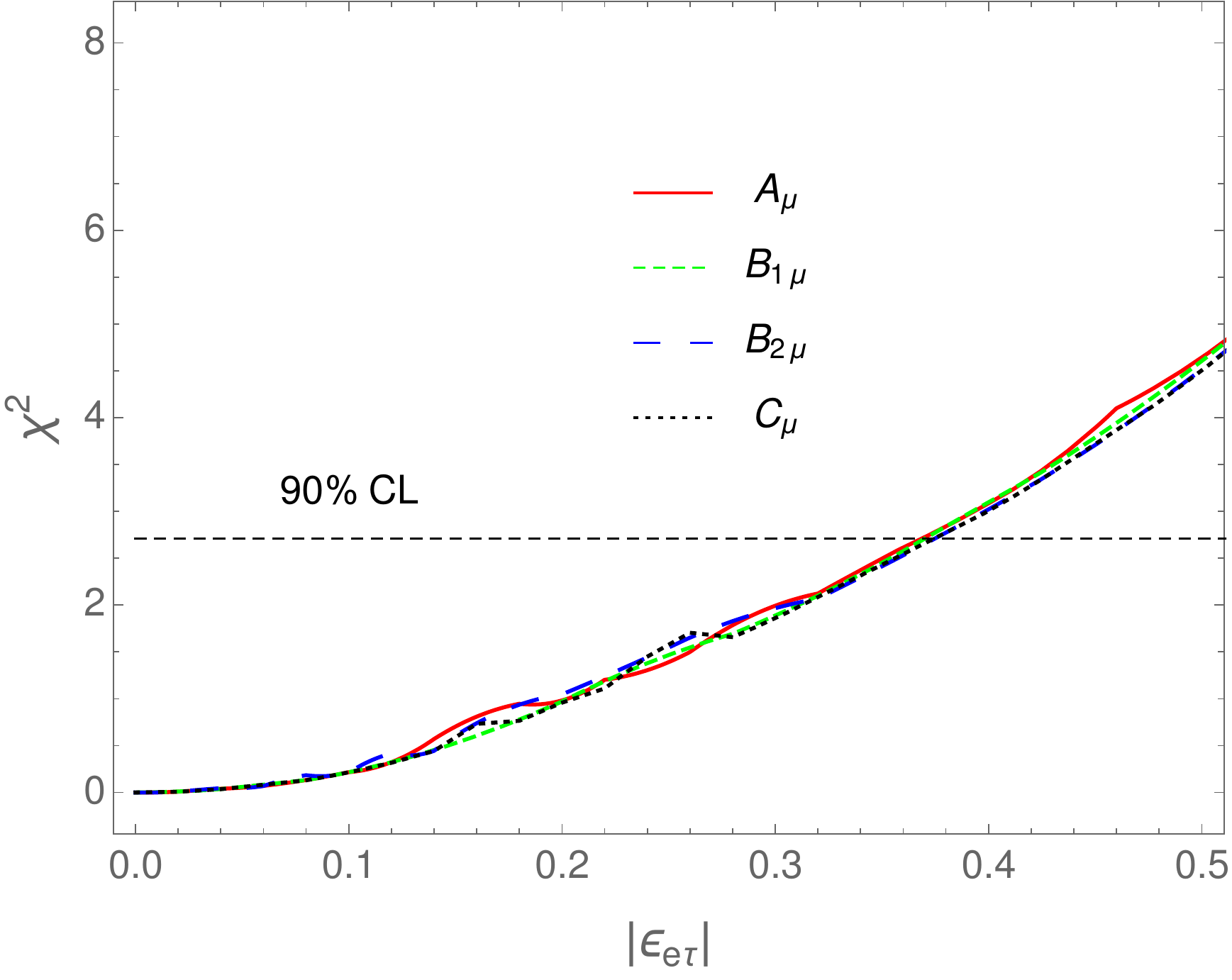}
\caption{\it \label{fig:chi22mu} The same as Fig.(\ref{fig:chi2mu}) but for the NSI parameters 
$\epsilon_{ee}$ (left panel), $|\epsilon_{e\mu}|$ (central panel) and $|\epsilon_{e\tau}|$ (right panel).
Vertical (green) bands represent the 90\% allowed regions summarized in Tab.(\ref{tab:limits}).}
\end{center}
\end{figure}

For the case-$II$, the dependence of $P(\nu_\mu \to \nu_\mu)$ on the NSI parameters suggests that a change in the systematics affects more $\epsilon_{\mu\mu}$ and $\epsilon_{\mu\tau}$ than the others;
this is clearly visible in Fig.(\ref{fig:chi2mu}), where we reported the $\chi^2$ function for $\epsilon_{\mu\mu}$ (left panel) and $|\epsilon_{\mu\tau}|$ (right panel), and in 
Fig.(\ref{fig:chi22mu}) where  $\epsilon_{ee}$ (left panel), $|\epsilon_{e\mu}|$ (central panel) and $|\epsilon_{e\tau}|$ (right panel) have been considered.

The trend of having much better bounds in the cases ($A_\mu$, $B_{1\mu}$) than ($B_{1\mu}$, $C_\mu$) is confirmed by our numerical results although, 
if compared to the case-$I$, there is a 
more clear separation among the curves in the same pairs. This is due to the fact that the muon disappearance
is now the channel under discussion, which provides a much better statistics than the appearance one. 
Going into details, we observe that the whole allowed range of $\epsilon_{\mu\mu}$ improves by $\sim$26\% from case $C_\mu$ to $A_\mu$ while that of 
$|\epsilon_{\mu\tau}|$ by roughly 15\%, see Fig.(\ref{fig:chi2mu}), thus making extremely relevant a reduction of the shape uncertainty for the 
$\nu_\mu$ signal.
No significant changes (below 3\%) are observed for the other moduli, see Fig.(\ref{fig:chi22mu}), not even for $\epsilon_{ee}$ for which the 
$\epsilon^3$-dependence in $P(\nu_\mu \to \nu_\mu)$ cannot be compensated and made comparable with the $\epsilon$-dependence of 
$\epsilon_{\mu\mu}$ and $\epsilon_{\mu\tau}$ by ${\cal O}(1)$ moduli.

In Tab.(\ref{tab:summary2}) we report the obtained 90\% CL bounds also for the case-$II$. The most relevant feature is that, like for case-$I$, 
more stringent bounds by a factor of $\sim$3 can be 
set on $|\epsilon_{e\mu}|$ and, to a lesser extent, on $|\epsilon_{e\tau}|$, while it is clear that DUNE cannot improve the 
current limits for all other parameters, even in the most aggressive scenarios for the systematic uncertainties. 
\begin{table}[h!]
\begin{center}
\renewcommand\arraystretch{1.5}
\begin{tabular}{c  c  c  c  c}
\toprule
\midrule
\bf parameter & $A_\mu$& $B_{1\mu}$ & $B_{2\mu}$ & $C_\mu$ \\
\midrule
$ \epsilon_{ee}$ & [-1.84, 2.04] & [-1.85, 2.04] & [-1.89, 2.10] & [-1.89, 2.10] \\ 
$|\epsilon_{e\mu}|$   & [0, 0.127]& [0, 0.130]& [0, 0.131]& [0, 0.131]\\ 
$|\epsilon_{e\tau}|$   & [0, 0.37] &[0, 0.37] &[0, 0.37] &[0, 0.37] \\ 
$\epsilon_{\mu\mu}$   & [-0.37, 0.46] & [-0.38, 0.44]& [-0.42, 0.50]& [-0.43, 0.51]\\ 
$|\epsilon_{\mu\tau}|$  & [0, 0.28]& [0, 0.29]& [0, 0.31]& [0, 0.33]\\ 
\midrule
\bottomrule
\end{tabular}
\caption{\it  
Case-$II$: summary of the 90 \% CL bounds that DUNE can set on the moduli of the NSI parameters  as the systematic uncertainties vary from $A_\mu$ to $C_\mu$.
}
\label{tab:summary2}
\end{center}
\end{table}

\section{Effects of NSI parameters on the extraction of standard parameters}
\label{sect:effect}
It is  interesting to estimate how the presence of matter NSI can worsen the sensitivity to the octant of $\theta_{23}$, 
to mass ordering and to CP violation at DUNE. To make things simpler, we limit ourselves to two extreme  scenarios, built 
from the two cases analyzed above: the $(A,A_\mu)$ scenario (OPTimistic), which means that the shape normalization error is 2\% 
and the absolute normalization error is 2.5\% for both $\nu_e$ and $\nu_\mu$ signals, and 
the $(C,C_\mu)$ scenario (PESsimistic) in which the shape normalization error is 7\% 
and the absolute normalization error is 5\%, again for $\nu_e$ and $\nu_\mu$ signals.
In this way we are confident that our results will include a large class of intermediate assumptions on systematic errors. 
In both situations, the errors and correlations among the several sources of backgrounds are the same as above.
Unless stated otherwise, in all plots were NSI parameters are taken into account, they have been marginalized over assuming the bounds reported in Tab.(\ref{tab:limits}).

\subsection*{Determination of the octant of $\theta_{23}$}
\begin{figure}[h!]
\begin{center}
  \includegraphics[scale=0.6]{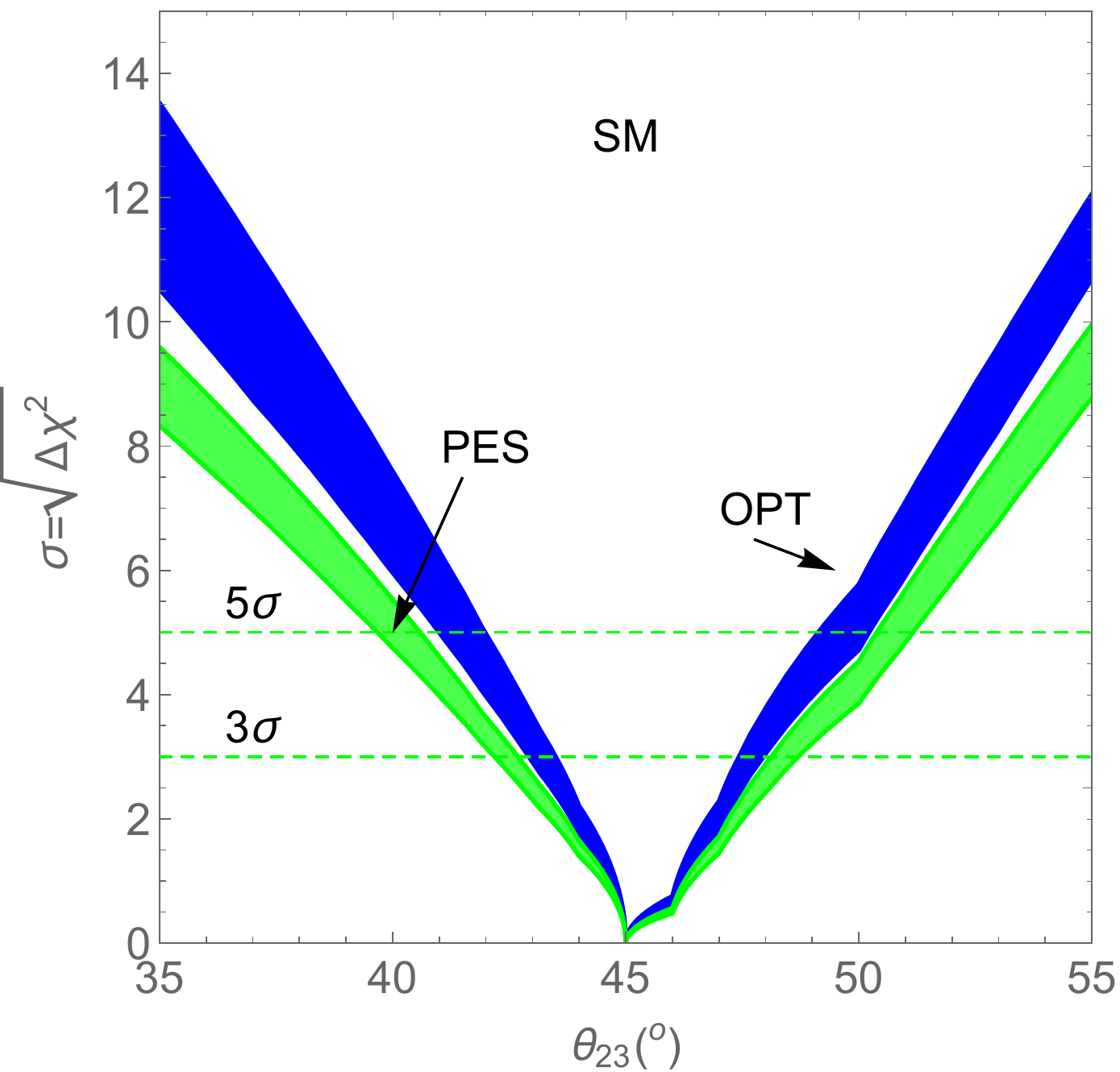} \includegraphics[scale=0.585]{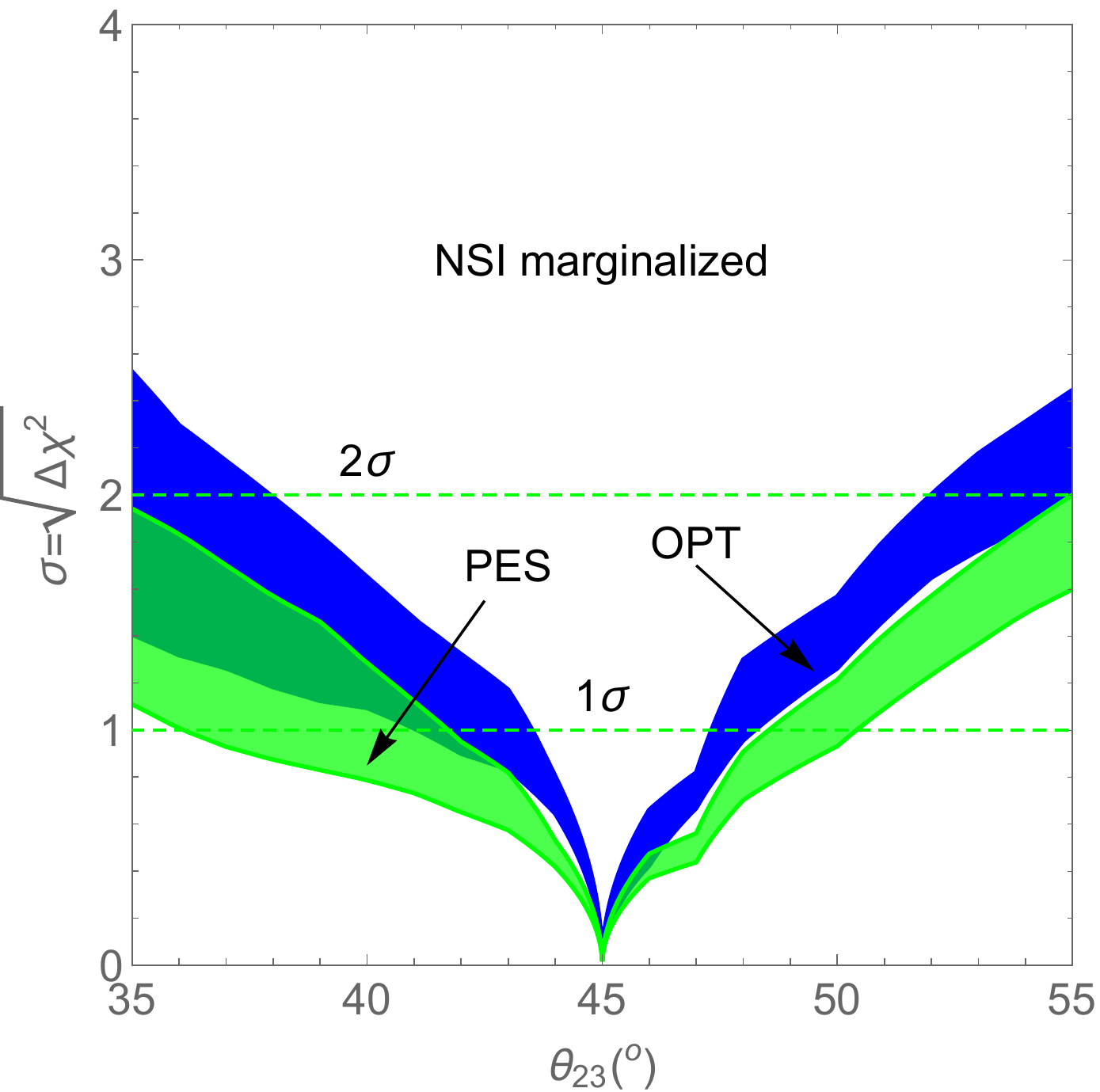}
\caption{\it \label{fig:octant}  Determination of the octant of $\theta_{23}$ as a function of the true value of the atmospheric angle, for NO. 
Left panel: results obtained when all NSI parameters are fixed and set to zero. Right panel: NSI parameters are marginalized over (notice the different vertical scale).
Dashed horizontal lines indicate the 3$\sigma$ and 5$\sigma$ CL in the left panel and 1$\sigma$ and 2$\sigma$ CL in the right one (pay attention to the different vertical scale).
The  bands represent the range in sensitivity due to potential
variations in the true value of $\delta_{CP}$: in light green the results obtained in the PES scenario and in blue those in the OPT case.
The parameters not shown are marginalized over. }
\end{center}
\end{figure}

The uncertainty in the determination of whether $\theta_{23}$ is larger or smaller than maximal mixing stems from the fact that 
its measurements are mainly due to the $\nu_\mu$ disappearance channel  which depends on $\sin^2 2\theta_{23}$. In this respect, the fact 
the DUNE can have simultaneous access to both $\nu_\mu$ disappearance and $\nu_e$ appearance channels provides an useful synergy to probe the 
octant hypothesis.
To estimate the sensitivity, we adopt the following metric \cite{DeRomeri:2016qwo}:
\begin{eqnarray}
 \Delta \chi^2 = \chi^2 (\pi/2 - \theta_{23}^{true}) -  \chi^2 ( \theta_{23}^{true})\,,
\end{eqnarray}
without imposing any priors on the atmospheric angle.
Our results (for NO only) are reported in Fig.(\ref{fig:octant}).
In the right panel  we show the situation when all NSI parameters are fixed and set to zero, thus this corresponds to pure SM (all NSI vanishing) results.
The dashed horizontal lines indicate the 3$\sigma$ and 5$\sigma$ CL; the area in the light green  region represents the range in sensitivity due to potential
variations in the true value of $\delta_{CP}$ in the case of the PES scenario while the region in blue is obtained for the OPT case.
We clearly see that DUNE can determine the octant at high confidence level as soon as $\theta_{23}$ is away from maximal mixing  by $\sim 3^\circ$
for the OPT case and by $\sim 4^\circ$ for the PES case; to be more precise, in the most favorable cases, a 5$\sigma$ discovery of the octant would be guaranteed outside 
the intervals $\theta_{23}\in [42.0^\circ,49.1^\circ]$
for the OPT assumption and $\theta_{23}\in [40.7^\circ,50.2^\circ]$ for PES.
This also means that for the $\theta_{23}$ central value reported in Tab.(\ref{tab:limits}) the significance with which DUNE can resolve the 
octant is scarse \cite{Nath:2015kjg}. Notice also that the difference between the PES and OPT cases is more evident for smaller values of 
$\sin^2 2\theta_{23}$, where $\sqrt{\Delta \chi^2}$ can be as large as four units.
The situation gets sensibly worse when marginalization over the NSI parameters is taken into account, as we can see in the right panel of Fig.(\ref{fig:octant}).
In particular, the discovery of the octant drops below 2$\sigma$ in the PES case even  for values of $\theta_{23}$ at the extremes of the range analyzed here  
 while a 2$\sigma$ sensitivity can only be reached in the OPT case and for mixing angles outside the interval 
$\theta_{23}\in [38^\circ,52^\circ]$, as one would have expected \cite{Agarwalla:2016fkh},
and for favorable values of the standard CP phase.

\subsection*{CP violation sensitivity}
In the standard framework of three neutrino oscillation, a signal indicating violation of the CP symmetry in the lepton sector will be observable if the true value of $\delta_{CP}$
 is sufficiently different from the CP conserving values $\delta_{CP}=0$ and $\delta_{CP}=\pi$. In order to estimate the capability of DUNE to determine leptonic CP violation,
we make use of the following indicator:
\begin{eqnarray}
\Delta \chi^2_{CPV} = Min\left[ \Delta \chi^2_{CP}(\delta_{CP}^{test} =0) ,
\Delta \chi^2_{CP}(\delta_{CP}^{test} =\pi) \right]\,,
\end{eqnarray}
where $\Delta \chi^2_{CP} = \chi^2_{\delta_{CP}^{test}} - \chi^2_{\delta_{CP}^{true}}$.
Given the fact that the smallest the value of $\sin^2 \theta_{23}$ the largest the sensitivity to CP violation is, we generally expect a smaller significance 
with respect to the results quoted in \cite{Acciarri:2015uup} even in the case of the SM only, as our $\sin^2 \theta_{23} \sim 0.59$ while in \cite{Acciarri:2015uup} they 
used $\sin^2 \theta_{23} = 0.45$ (and similar values for the other mixing angles).
In the fit procedure we marginalized over all undisplayed parameters; the true value of $\theta_{23}$ is set in the second octant, according to the 
central values quoted in Tab.(\ref{bestfit}).

\begin{figure}[h!]
\begin{center}
  \includegraphics[scale=0.6]{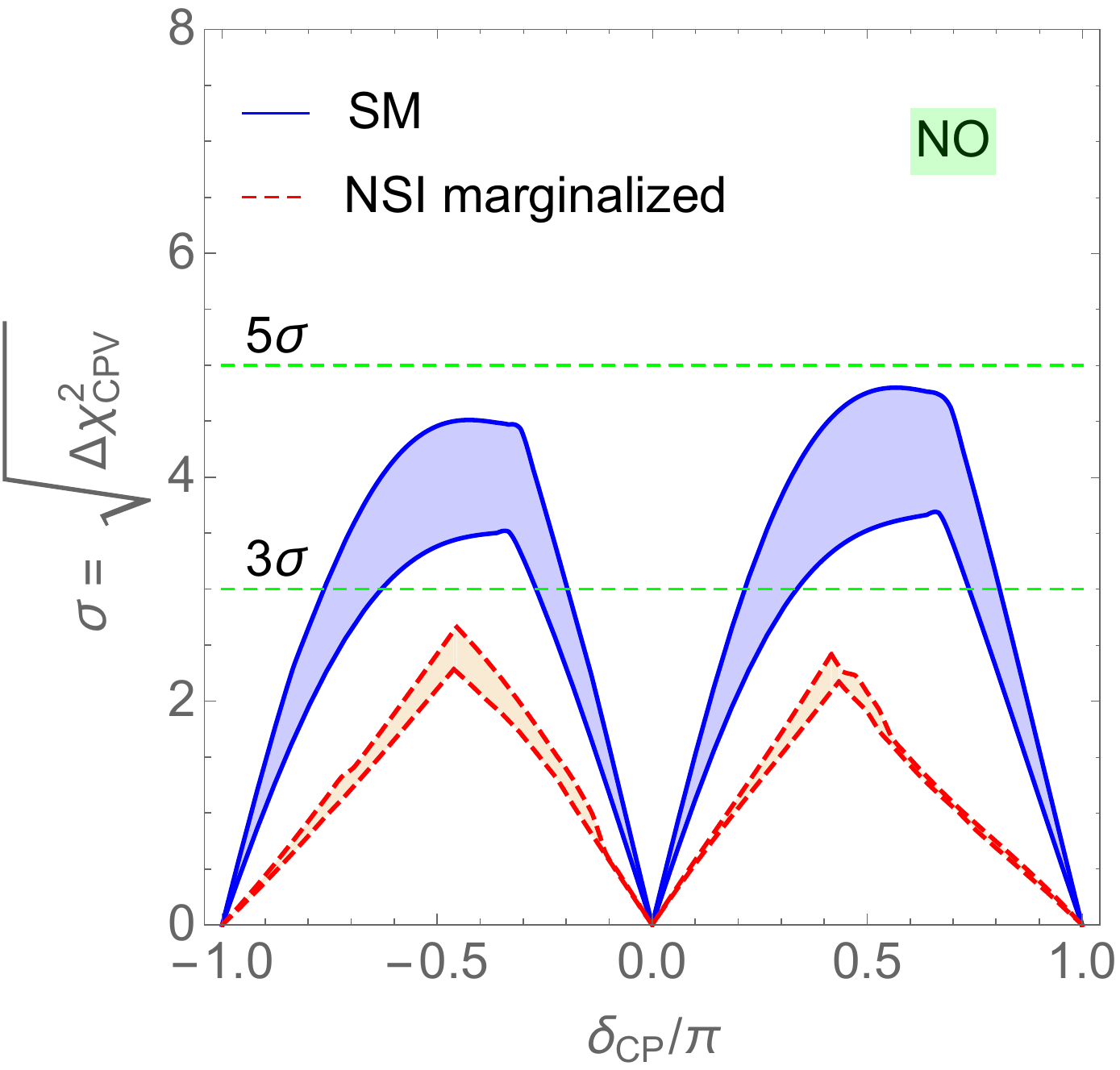} \includegraphics[scale=0.6]{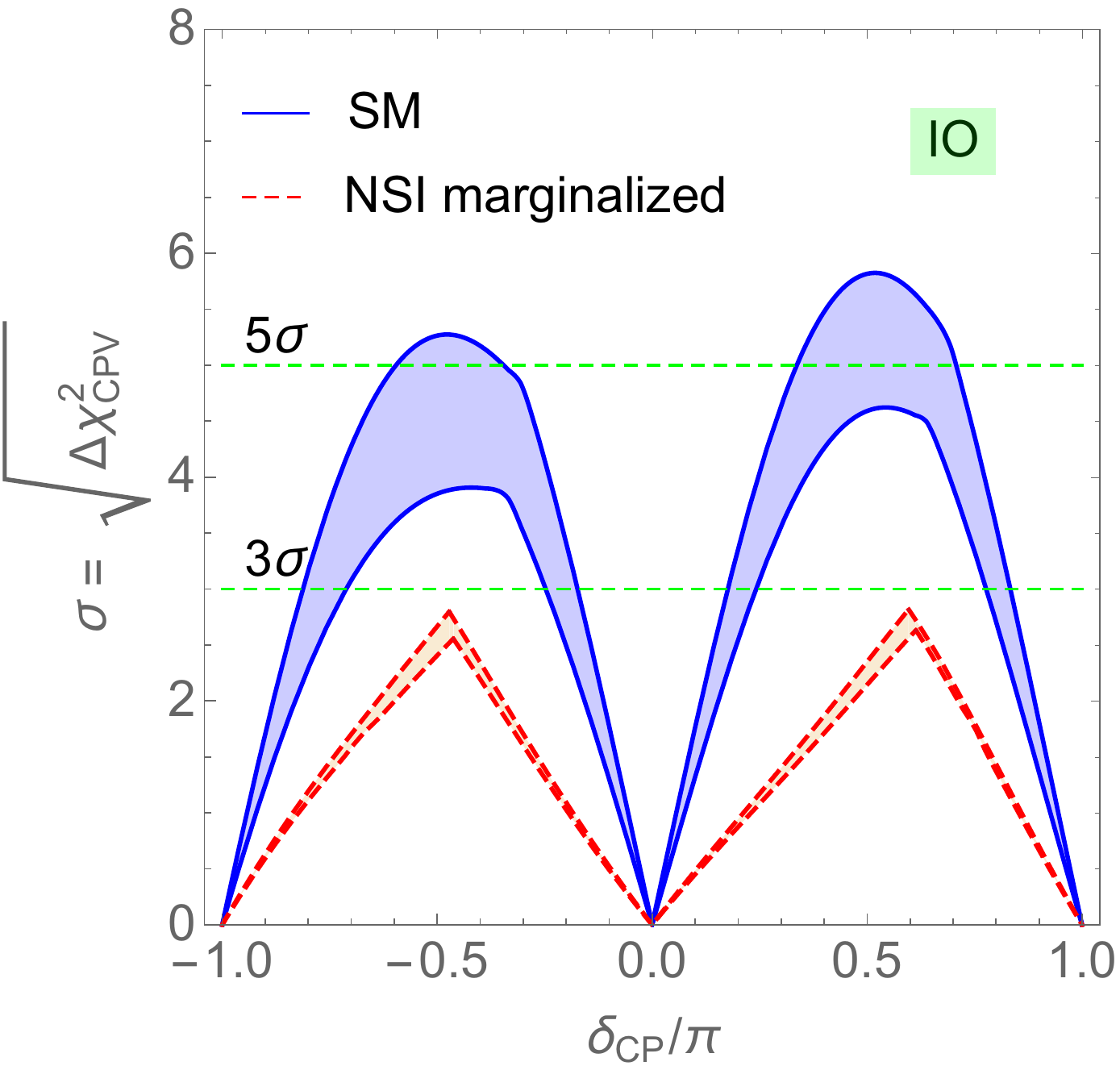} 
\caption{\it \label{fig:cpv}  CP discovery potential  as a function of the true value of the leptonic CP phase for NO (left panel) and IO (right panel). 
Dashed horizontal lines indicate the 3$\sigma$ and 5$\sigma$ CL.
The bands represent the range in sensitivity obtained under the two different assumptions for the systematics: upper curves in each bands
are generated in the OPT case, the lower ones in the PES case. Blue curves/bands are for SM, red/dashed when NSI are marginalized over.}
\end{center}
\end{figure}

Our results are reported in Fig.(\ref{fig:cpv}) where we displayed the 
CP discovery potential  as a function of the true value of the leptonic CP phase, for both NO (left panel) and IO (right panel). 
The bands represents the range in sensitivity obtained under the two different assumptions for the systematics, with the implicit meaning that the upper curves in each bands
correspond to the OPT case while the lower ones to the PES case. The results for the SM are in blue/solid line, those considering the marginalization over the NSI parameters in red/dashed.

In the SM case, the DUNE setup adopted in this paper is enough to guarantee a 5$\sigma$ discovery potential for the leptonic phase only for the inverted 
ordering of the neutrino mass eigenstates, in $\sim$31\% of the cases in the OPT scenario; if PES systematics are considered, only a 3$\sigma$ discovery potential
can be reached in $\sim$50\% of the true $\delta_{CP}$.
For NO and for the OPT case,  an 
encouraging 4$\sigma$ can be reached in the $\sim$39\% of the $\delta_{CP}$ values, roughly the same percentage obtainable in the PES case at 3$\sigma$. 
The marginalization over the NSI parameters 
strongly reduces the CP discovery potential \cite{Masud:2016bvp} such that not even 3$\sigma$ discovery can be claimed.

In this context the dependence on the choice of systematics is quite relevant for both SM and NSI-marginalized cases: the sensitivities for the OPT and PES assumptions differ by $\sim$20\% and 
$\sim$15\% for $|\delta_{CP}|\sim \pi/2$, respectively, almost independently on the considered hierarchy.

\subsection*{Sensitivity to mass hierarchy}
\begin{figure}[h!]
\begin{center}
  \includegraphics[scale=0.45]{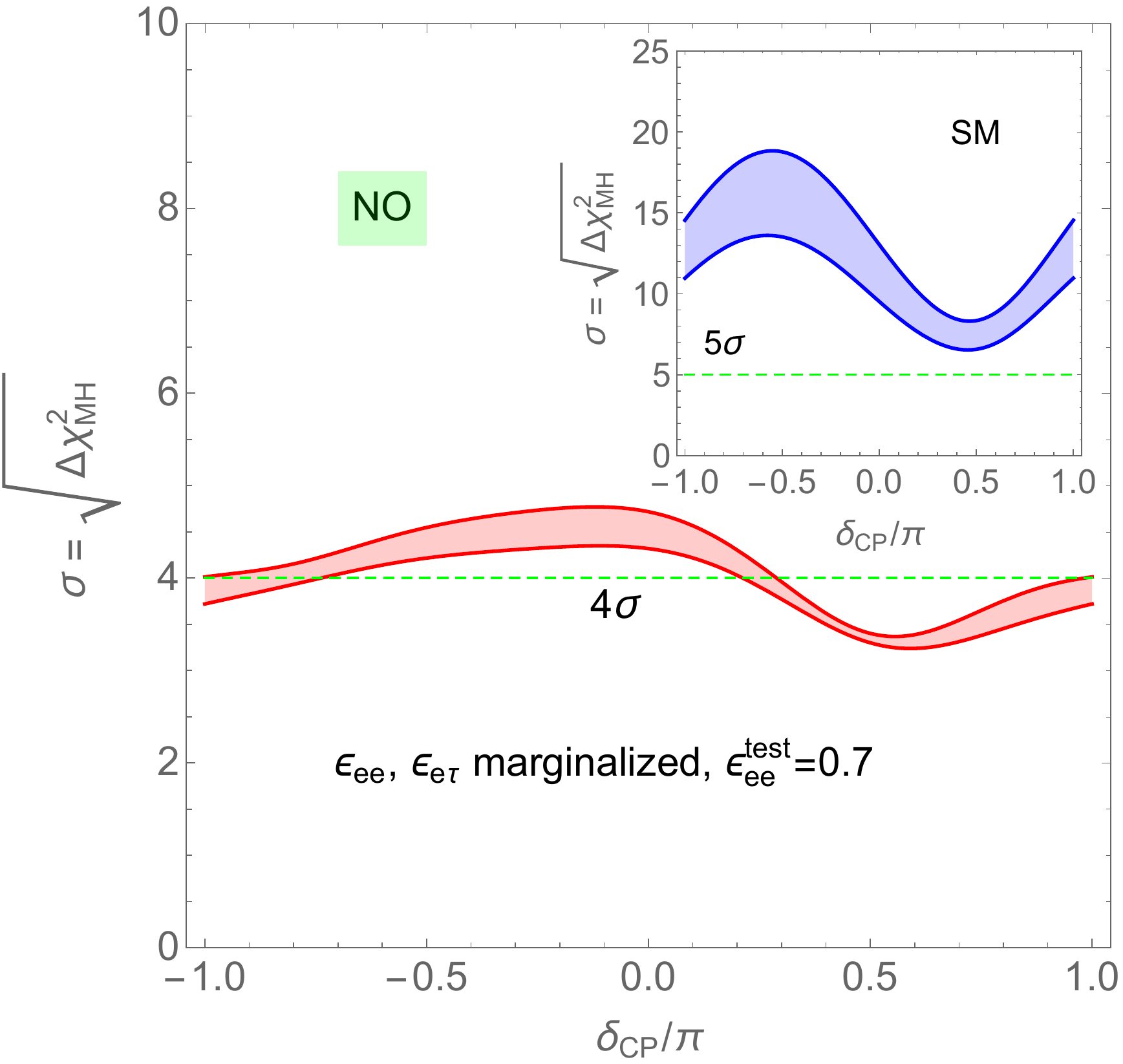} \includegraphics[scale=0.45]{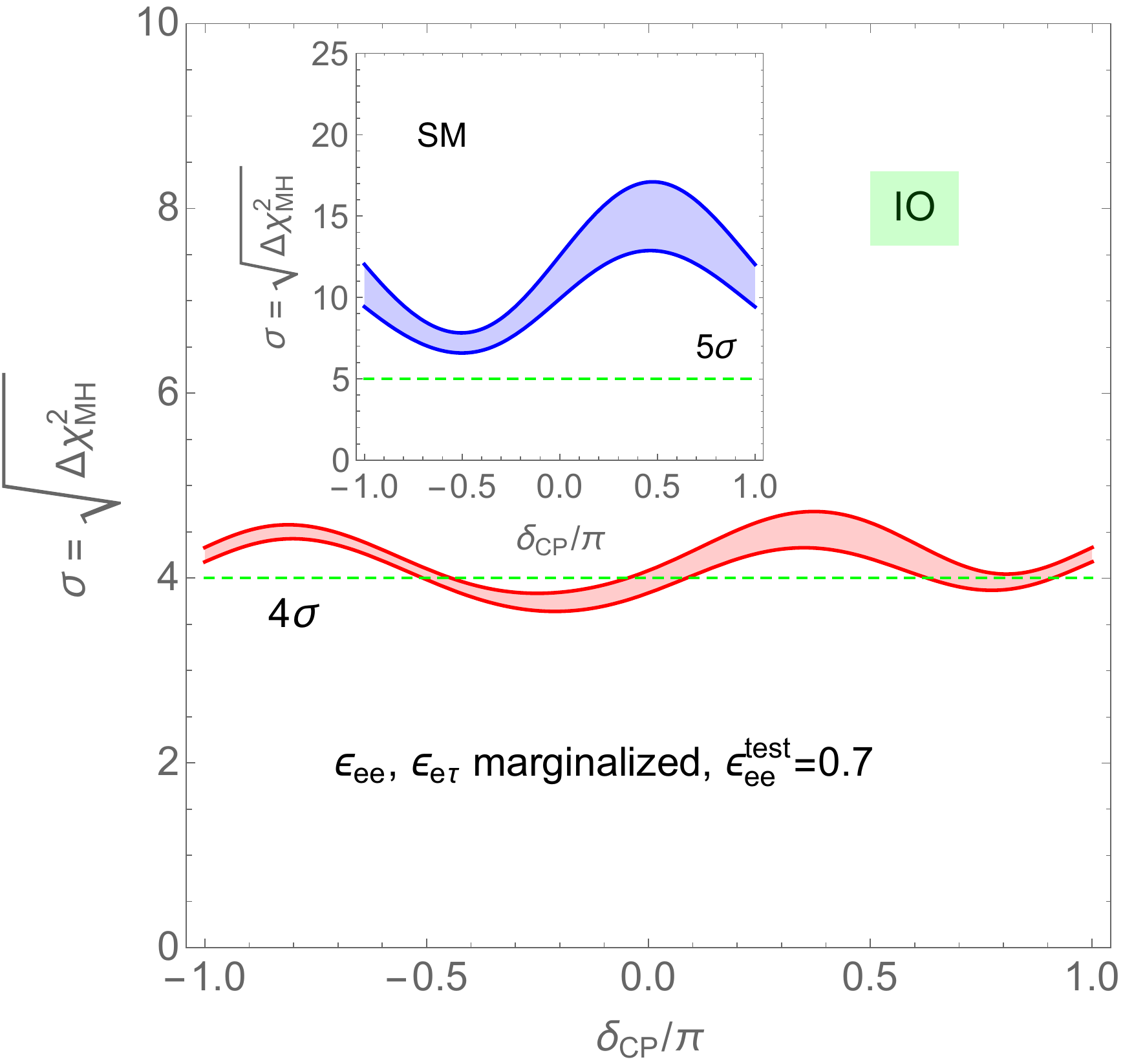} 
\caption{\it \label{fig:mass}  Significance with which the mass hierarchy can be determined as a function of the true value of the CP phase $\delta_{CP}$. 
Left panel: results obtained assuming that NO is the true mass ordering. Right panel: results obtained assuming IO as the true mass ordering.
The  bands represent the range in sensitivity due to the different assumptions on the systematics. In the boxes the results for the standard three neutrino oscillations are reported.}
\end{center}
\end{figure}

The long-baseline and high neutrino energies of the DUNE flux are particularly suited to explore the sensitivity to mass hierarchy, that is to determine the 
sign of the atmospheric mass difference $\Delta m^2_{31}$.
As it is well known \cite{Coloma:2016gei}-\cite{Deepthi:2016erc}, the marginalization over all NSI parameters produces a loss in the sensitivity to mass hierarchy (both for NO and IO); 
in our numerical simulations, this barely reaches values larger than $\sqrt{\Delta \chi^2} \sim 1.4$ when
$\delta_{CP} \sim \pi/2$. This conclusion can be traced back to the intrinsic degeneracy involving $\epsilon_{ee}$ and $\delta_{CP}$. Since the primary goal of this paper is 
to illustrate the effects of systematics, we decided to marginalize $\epsilon_{ee}$ in the positive interval reported in Tab.(\ref{tab:limits}), with a true
value $\epsilon_{ee}=0.7$ and to keep fixed all other NSI parameters but $\epsilon_{e\tau}$, whose correlation with $\epsilon_{ee}$ has been shown to be important 
in decreasing the sensitivity of long-baseline experiments to several physical quantities \cite{Liao:2016hsa}. 
We quantify the experimental sensitivity to the mass hierarchy using\footnote{As shown in Refs.\cite{Qian:2012zn,Blennow:2013oma}, 
the $\Delta \chi^2$ metric does not follow the $\chi^2$ function; it is used here as a representative of the mean 
or the most likely value of the true $\Delta \chi^2$ that would have been obtained in an ensemble of experiments.}:
\begin{eqnarray}
 \Delta \chi^2_{MH} &=& \chi^2_{IO} - \chi^2_{NO} \qquad {\rm true\; normal\; ordering} \nonumber\,, \\ 
 \Delta \chi^2_{MH} &=& \chi^2_{NO} - \chi^2_{IO} \qquad {\rm true\; inverted\; ordering}\nonumber \,.
\end{eqnarray}
Our results are reported in Fig.(\ref{fig:mass}) where we show $\sqrt{\Delta \chi^2_{MH}}$ as a function of the true value of the CP phase $\delta_{CP}$, 
for the two cases where the true ordering is the NO (left panel) or the IO one (right panel). The plots in the boxes depict the same variable 
$\sqrt{\Delta \chi^2_{MH}}$ computed in the SM. The  bands represent the range in sensitivity due to the different assumptions on the systematics, 
with the meaning that the largest $\sqrt{\Delta \chi^2_{MH}}$ corresponds to the OPT case.
Within our setups, the mass ordering in the SM case can be determined at $\sqrt{\Delta \chi^2_{MH}} = 5$ for every value of the true $\delta_{CP}$ for both NO and IO while the same 
is not true when the NSI marginalization is taken into account: in this case, the mass hierarchy discovery potential is larger than $\sqrt{\Delta \chi^2_{MH}} = 3$ for every $\delta_{CP}$
and occasionally also larger than 4. 
While in the standard case (no NSI) the systematics change the sensitivity by 20 to 28\% (for both assumptions on the true neutrino mass ordering), 
when NSI are taken into account the hypothesis on the systematic effects change the results by a $\sim$12\% on average.

\section{Conclusions}
\label{sect:conc}
In this paper we tried to characterize the role of different assumptions on shape and normalization uncertainties for the 
$\nu_\mu$ and $\nu_e$ signals at the DUNE facilities in the determination of the bounds on the NSI parameters. 
The main results of our study can be summarized in the following points:
\begin{itemize}
\item shape uncertainty is much more relevant than the signal normalization uncertainty in setting the bounds on NSI parameters;
\item for $|\epsilon_{e\mu}|$, reasonable assumptions on the systematics (for both appearance and disappearance channels) 
are enough to lower by a factor of $\sim$3 the existing upper limits;
\item on the other hand, for  $|\epsilon_{e\tau}|$ only a 30-40\% reduction of the current upper limits can be reached in DUNE, depending on
whether the assumptions of case-$I$ or $II$ are taken into account, respectively;
\item $\epsilon_{\mu\mu}$ and $|\epsilon_{\mu\tau}|$ strongly benefit of smaller systematic uncertainties in the $\nu_\mu$ sector but even in the most favorable cases, $A$ and $A_\mu$,
the current limits cannot be ameliorated;
\item for $\epsilon_{ee}$, the configuration of DUNE used in this paper helps in disfavoring negative values, independently on the assumptions on systematics. 
\end{itemize}
In addition to the previous considerations, we have studied the impact of the marginalization on the NSI parameters on 
the sensitivity to the octant of $\theta_{23}$, to mass ordering and to CP violation at DUNE. In the case of the octant of $\theta_{23}$, 
a 2$\sigma$ sensitivity in the NSI-marginalized case can be reached for very optimistic assumptions on the systematics 
and for mixing angles outside the interval 
$\theta_{23}\in [38^\circ,52^\circ]$ and for favorable values of the standard CP phase.
For the sensitivity to CP violation, the choice of systematics can alter the results at $|\delta_{CP}|\sim \pi/2$ by 15-20\%, where the largest impact is seen in the SM case; 
in this respect not much difference has been found for the different hierarchies. Finally, we have observed that the choice of the systematics is not crucial to get  a sensitivity 
$\sqrt{\Delta \chi^2_{MH}}$ above 5$\sigma$ in the SM case and well above 3$\sigma$ in the NSI-marginalized case for every value of $\delta_{CP}$, even though the effects 
of changing from the PES to the OPT assumptions are well above 20\% and 10\%, respectively.

\section*{Acknowledgments}
I am grateful to Elizabeth Worcester for illuminating discussions on the use of the GLoBES software for implementing the bin-by-bin systematic 
uncertainties and to Davide Sgalaberna, Stefania Bordoni, Leigh Whitehead and Paola Sala for useful suggestions on reproducing the DUNE CDR results using GLoBES. 

\bibliographystyle{MLA}

\end{document}